\documentclass[reprint, aps, onecolumn, amsmath, amssymb,footinbib]{revtex4-2}

\usepackage[margin=2.6cm]{geometry}

\usepackage{hyperref}
\hypersetup{
    colorlinks = true,
    urlcolor   = blue,
    citecolor  = black,
}
\hypersetup{colorlinks,linkcolor={blue},citecolor={blue},urlcolor={blue}} 
\usepackage{graphicx}
\usepackage{dcolumn}
\usepackage{bm}
\usepackage[separate-uncertainty = true,multi-part-units=single]{siunitx}
\usepackage{mathtools}
\usepackage{gensymb}
\usepackage{amsmath}
\usepackage{physics}
\usepackage{stmaryrd}
\usepackage[nameinlink,noabbrev]{cleveref}
\crefformat{section}{\S#2#1#3}%
\crefformat{subsection}{\S#2#1#3}
\crefformat{subsubsection}{\S#2#1#3}
\crefformat{figure}{#2Fig.~~#1#3}
\crefmultiformat{figure}{Figs.~#2#1#3}{ and~#2#1#3}{, #2#1#3}{ and~#2#1#3}
\usepackage{amssymb}
\usepackage[svgnames]{xcolor}
\colorlet{NewText}{black}
\usepackage{soul}
\setstcolor{red}
\usepackage{wasysym}

\usepackage{natbib}
\usepackage{xcolor}
\usepackage{soul}
\newcommand{\be}{\begin{equation}}
\newcommand{\ee}{\end{equation}}

\begin{document}

\preprint{APS/123-QED}

\title{Light-scattering reconstruction of transparent shapes using neural networks}

\author{Tymoteusz Miara$^1$}\thanks{Current address: School of Biological and Behavioural Sciences, Queen Mary University of London, Mile End Road, London E1 4NS, UK}\email{tymoteuszmiara@gmail.com}
\author{Draga Pihler-Puzović$^1$}
\author{Matthias Heil$^2$}
\author{Anne Juel$^1$} \email{anne.juel@manchester.ac.uk}
\affiliation{$^1$Department of Physics \& Astronomy, School of Natural Sciences,  University of Manchester,  Oxford Road,  Manchester M13 9PL, UK}
\affiliation{$^2$Department of Mathematics, School of Natural Sciences,  University of Manchester,  Oxford Road,  Manchester M13 9PL, UK}

\date{\today}

\begin{abstract}
The accurate characterisation of the 3D deformations of slender fibres and thin sheets in flow, is a key experimental challenge in the study of particle-laden flows. We propose a high-resolution, single-camera method to visualise non-intrusively the shape of a transparent crumpled sheet, as it translates, rotates and deforms. We perform periodic scans of the crumpled shape by illuminating it with a sequence of stacked light sheets at a rate much faster than its deformation and image the scattered light signal in a plane near-orthogonal to the plane of lighting. Processing of the data using a pinhole camera model yields a noisy spatio-temporal dataset of the strongly deformed time-evolving surface of the sheet, which we reconstruct in 3D using a neural autoencoder. We validate the robustness of the shape reconstruction algorithm to noise using synthetic data sets, and demonstrate the accurate reconstruction of laboratory sedimentation experiments with elastic disks. We find that the inclusion of isometricity-enforcing penalties into the cost function of the autoencoder enables us to robustly reconstruct highly folded shapes, where different regions of the sheet overlap.
\end{abstract}

\maketitle

\section{Introduction}

Deformation of slender fibres and elastic sheets 
in a fluid flow is a subject of active research, with examples across scales ranging from flapping flags~\cite{TAVALLAEINEJAD2021103199} to microplastic dispersion~\cite{annurev:/content/journals/10.1146/annurev-fluid-120423-012604}, the handling of actin filaments~\cite{doi:10.1073/pnas.1805399115}, and processing of graphene sheets in liquid environments~\cite{Silmore2021}. These particles are prone to adopting complex shapes under load because they have a small resistance to out-of-plane elastic bending deformation. Recent studies have shown that elastic fibres and sheets can exhibit intricate dynamics of deformation and reorientation in low Reynolds number sedimentation and shear flows \cite{FlexibleFibreExperimental2018, Graham2024, Silmore2021}, while the environmental and health threat posed by microplastic contamination is driving interest in understanding the dispersion of elastic fibres and thin deformable sheets in turbulent flow at high Reynolds number~\cite{Sugathapala2025}. 
However, visualisation of morphing shapes and dynamics of elastic sheets in flow is a significant experimental challenge.

Methods of 3D reconstruction typically rely on stereoscopic imaging, where two (or more) cameras view an object from different angles \cite{3Dvision}. A single camera may suffice in digital image correlation of speckle patterns embedded at the surface of the material \cite{Haolin} or laser triangulation, which involves the analysis of the distortion of a projected laser line~\cite{PhysRevLett.111.154501,  Ben-Shachar_Brumley_Hogg_Hinton_2025}. Quantitative synthetic Schlieren imaging methods based on fast Fourier demodulation have recently been extended from the capture of wavy fluid interfaces to quantifying the local elevation on a deformed elastic sheet, where instead of a random dot pattern, a checkerboard pattern is used as a backdrop to the refractive object \cite {Wildeman, Khatla}. Large deformations associated with the dynamical crumpling of a gel sheet have been accurately captured by seeding the transparent material with fluorescent particles. Repeated scanning of the shape with a projected laser line enables the fluorescent outlining of the shape and thus, direct observation of spatio-temporal crumpling dynamics \cite{Aharoni2010}.  However, all these methods rely on moderate surface slopes, such that the entire surface to be reconstructed is in direct vision of the camera. To resolve the inner layers of the crumpled material, it becomes necessary to resort to X-ray tomography \cite{Lin2009}.

Although single-camera methods are also commonly used for tracking elastic fibres in low Reynolds number flows~\cite{doi:10.1073/pnas.1805399115, Chakrabarti} and rigid fibres in turbulent flow~\cite{Voth2017}, they restrict knowledge of the fibre orientation to the imaging plane. For large-aspect-ratio elastic fibres, which easily deform into fully 3D configurations~\cite{Verhille2014}, at least three synchronised cameras are required to capture the evolving geometry and attaining sufficient spatial and temporal resolution to track complex deformations in homogeneous turbulence is difficult~\cite{Marchioli2026}. Using three synchronised high-speed cameras and a pinhole camera model to determine the position of a point in space from its projections onto the different images, Verhille \& Bartoli \cite{Verhille2016} apply shape-from-silhouette methods from computer vision \cite{Cheung2005} to reconstruct elastic fibres and disks. To process their data, they use the Convex Hull Volume method which only identifies the convex envelope of the object rather than its shape. The method successfully extends to elastic disks, which deform into U-bent disks when immersed in turbulent \cite{Verhille2022} \textcolor{NewText}{or vortical \cite{ibarra2023} flow}. Another powerful method for obtaining 3D particle orientations from single-camera measurements is digital holography, which reconstructs a 3D image of the object from the interference pattern between the object-diffracted beam and the reference light beam \cite{Voth2017}. However, two orthogonal cameras are still needed to resolve the object position in 3D. 

In this paper, we present an alternative low-cost method that uses a single camera to accurately reconstruct the evolving shape of a sedimenting elastic sheet from its folded initial configuration. Following Aharoni et al. \cite{Aharoni2010}, we employ transparent circular sheets (i.e., disks), but instead of embedding fluorescent particles in the sheet, we simply exploit Rayleigh scattering which illuminates the points of intersection between the deformed sheet and the scanning light.
The experimental setup for observing the sedimentation of an elastic disk, and for scanning its motion and deformation in top-view using a single-camera system, is described in Section \ref{S:Setup}.  
The subsequent step-by-step processing of these scans based on a pinhole camera model \cite{Verhille2016}, is discussed in sections~\ref{s:CameraOptics}--\ref{s:ImageProcessing}. This process yields a space-time representation of the deformed disk referred to as a \textit{hypercloud}.
By coupling the hypercloud with a neural autoencoder, which we show to be robust to unavoidable experimental noise, we reconstruct the disk shapes in 3D, even for highly folded configurations. 
The algorithm for obtaining a parametric representation of these shapes is presented in section \ref{s:Reconstruction}. Validation of the method using a synthetic dataset representative of the experimentally observed disk dynamics is provided in Section \ref{sect:validation}. Results are presented in section \ref{s:results} where time sequences of reconstructed, sedimenting, elastic disks are shown for different initial conditions ranging from U-bent disks to disks folded in four. Conclusions are presented in section \ref{s:conclusion}.

\section{Experimental setup and measurement method} \label{S:Setup}

\begin{figure}[t!]
    \centering
    \includegraphics[width=0.5\textwidth]{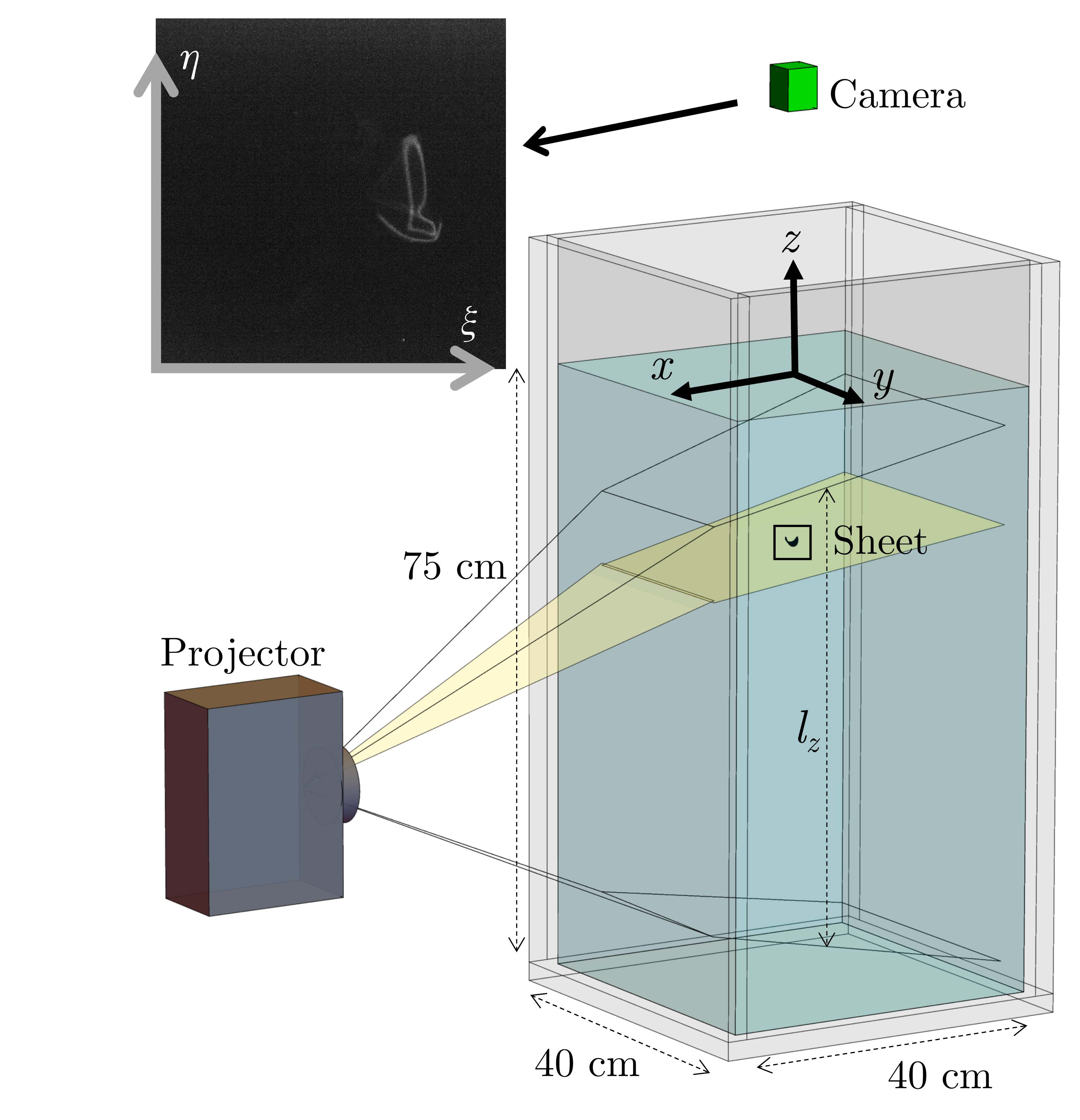}
    \caption{Schematic diagram of the experimental set-up (to scale), with a single plane of light shining through an elastic sheet. The transparent planes show the light sheets created by the first and last rows of pixels of the projector. Inset: a typical image captured by the top-view camera during a scan of a crumpled sheet.}
    \label{fig:CH2setup}
\end{figure}

We introduce our method in the context of the experimental setup shown in Figure~\ref{fig:CH2setup}. 
We study the sedimentation of a transparent elastic disk immersed in 1000 cSt silicone oil (Allcosil, J. Allcock \& Sons, with dynamic viscosity $\mu=1.02 \pm 0.01$\,Pa\,s) with density $\rho_\mathrm{f}=973\pm 0.5$~kg~m$^{-3}$. The disk has radius $R= 19.0\pm 0.2$~mm and is cut from  a PDMS sheet (Silex) of thickness $b=50\,\mu$m, which has density $\rho_\mathrm{s}=1073\pm 7$~kg~m$^{-3}$, Young's modulus $E=882\pm 1$~kPa and Poisson ratio $\nu\simeq 0.5$. 
The disk is initially crumpled and tends to relax towards a U-shape as it settles under gravity with a typical vertical velocity of 0.25~mm/s, in a liquid-filled tank with internal dimensions of $40\times40\times90\, {\rm cm^3}$; see Figure~\ref{fig:CH2setup}.
The approximate match of the refractive indices of the disk and the fluid means that the object is virtually invisible under ambient lighting. We illuminate the object using a high-definition projector (Optoma HD143X), which is positioned to the side of the tank, and casts oblique light sheets by displaying a single horizontal row of bright pixels. Rayleigh scattering at the intersection between the plane of light and the object renders the outline of the crumpled disk visible in this lighting cross-section. We capture the resulting light pattern in top-view with a JAI GO-5000M-USB camera fitted with a Kowa lens (LM16HC, RMA Electronics) in a $768\times768$~pixels$^2$ window shown in the inset image of Figure~\ref{fig:CH2setup}. The camera was positioned 25.9~cm above the liquid and its aperture was set to an f-number of 5.0 to ensure good sharpness across the depth of the illuminated region of $l_z=58.2$~cm (see Figure~\ref{fig:CH2setup}).  The absolute resolution at mid-depth of the tank is $166\,\mu$m/px  and the thickness of individual planes of light cast by the projector is $303\, \mu$m. 

\begin{figure}[t!] 
    \centering
    \includegraphics[width=\textwidth]{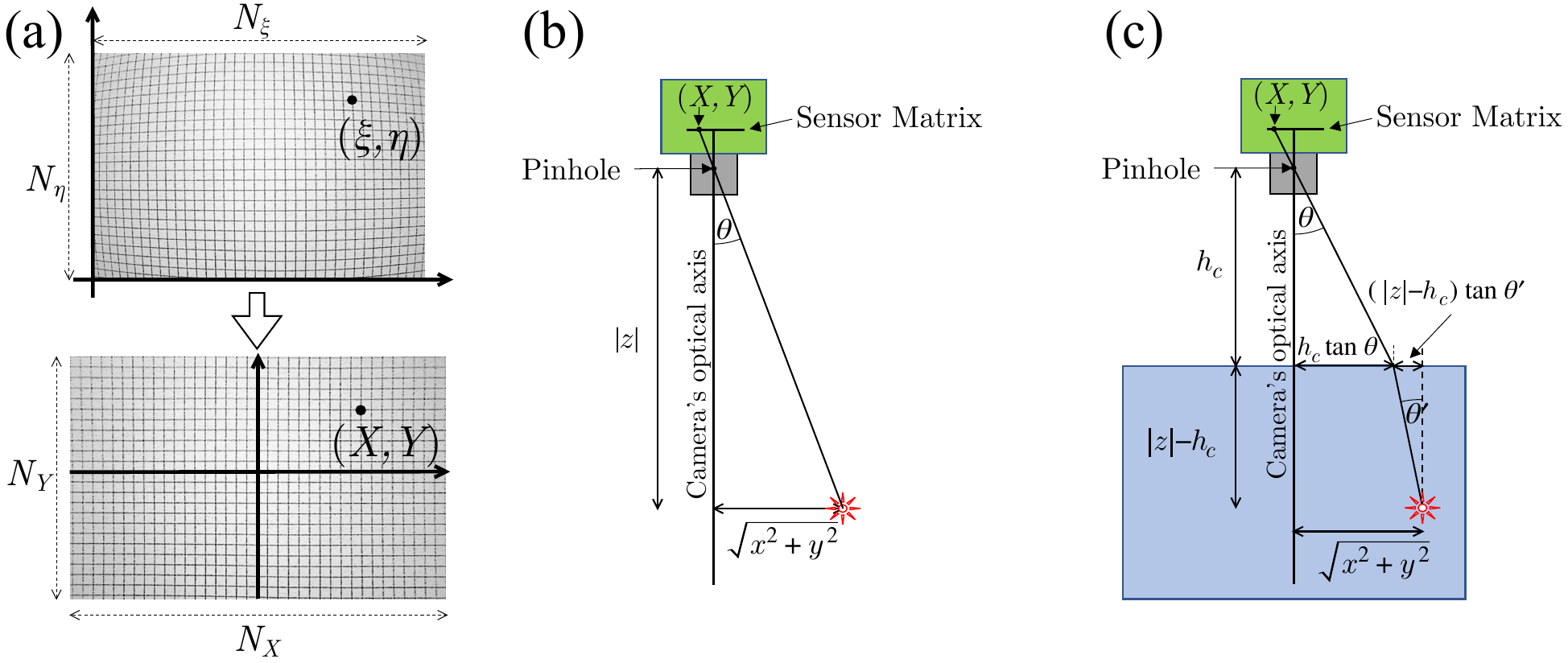}
    \caption{(a) Mapping of the distorted camera image due to lens aberration with pixel coordinates $(\xi, \eta)$ to the corrected coordinate $(X,Y)$. (b) Geometric optics of the top view camera capturing a luminous point in air. (c) Geometric optics of the top view camera capturing a luminous point inside a liquid.}
    \label{fig:CameraOptics}
\end{figure}

\subsection{Optics of the object-camera system}
\label{s:CameraOptics}

To track the shape evolution of the settling elastic disk, we need to identify the location in the tank of the features imaged by the camera as a function of time.
However, first we must relate the position of an illuminated point in the camera sensor coordinates system $(\xi, \eta)$
to its absolute position in the laboratory measured in a Cartesian coordinate system $(x,y,z)$, which we define so that $(x,y)$ are horizontal coordinates and the upward-pointing $z$-axis is aligned with the vertical optical axis of the camera; see Figure\;\ref{fig:CameraOptics}. \textcolor{NewText}{While such calibration could be done by tracking a small sphere moved by a set of three translation stages (see \cite{ibarra2023}), we simply rely on  solving for the refraction of light at the interface. To this end, it is essential for the camera's optical axis to be normal to the interface. We achieved this by ensuring that the camera images its own reflection exactly at the centre of the frame.}

We begin by correcting for image distortion due to lens aberration. For this purpose, we image a horizontal piece of graph paper shown in Figure~\ref{fig:CameraOptics}a and compute the corrected coordinates $(X,Y)$ from the camera pixel coordinates $(\xi,\eta)$ using the mapping
\begin{align}\label{eq:barrel}
    X(\xi,\eta)=\frac{\xi-N_\xi/2}{1+K_\mathrm{R} \left((\xi-N_\xi/2)^2+(\eta-N_\eta/2)^2\right)} \, ,  \\
    Y(\xi,\eta)=\frac{\eta-N_\eta/2}{1+K_\mathrm{R} \left((\xi-N_\xi/2)^2+(\eta-N_\eta/2)^2\right)} \, , 
\end{align}
where $N_\xi \times N_\eta$ are the number of pixels in the raw image captured by the camera. We determine the constant $K_\mathrm{R}$ by transforming the image until the corrected image is rectilinear (see bottom image in Figure~\ref{fig:CameraOptics}a); for our lens, $K_\mathrm{R}=13.5\times 10^{-9}$. \textcolor{NewText}{(Higher-order distortion parameters may be needed for some lenses, but were not necessary for our lens when viewing our region of interest)}.
The corrected image can be treated as originating from a pinhole camera with the sensor size $N_X \times N_Y=2 X(N_\xi,0)\times 2Y(0,N_\eta)$, which is larger than the original image if $K_\mathrm{R}<0$, as in Figure~\ref{fig:CameraOptics}a, and smaller if $K_\mathrm{R}>0$. The location of the effective pinhole is the point where all the incoming rays converge and is therefore a convenient choice for the origin of the laboratory's coordinate system. Provided the field of view (FOV) angle, $\theta_{\rm max}$, between the optical axis and the ray arriving at the corner of the image $(X,Y)=(N_X/2,N_Y/2)$ is known, the position of the pinhole can be determined relative to the opening of the lens by photographing a horizontal ruler positioned at a known distance from the lens, and calculating the distance to the pinhole, see Figure~\ref{fig:CameraOptics}b. If the FOV angle is not known, both the position of the pinhole and $\theta_{\rm max}$ can be determined by photographing the ruler from two known distances. For our system, $\theta_\mathrm{max}=55.8^\circ$ \textcolor{NewText}{and the effective pinhole is 29~mm behind the opening of the lens}.

We choose the $x$ and $y$-axes to be aligned with the horizontal and vertical axes of the camera's sensor matrix, and hence with the $X$ and $Y$ axes. From this alignment, we write down the first relation between the observed coordinates and the absolute position of the luminous point shown with a red marker in Figure~\ref{fig:CameraOptics}b,
\begin{equation} \label{eq:xyratio}
    \frac{x}{y}=\frac{X}{Y}.
\end{equation}

The second relation uses the fact that, in a pinhole camera, the rays pass through the pinhole in straight lines, thus the right-angled triangle formed by the pinhole, the luminous point and the optical axis is similar to the triangle between the pinhole, the $(X,Y)$ point and the centre of the sensor matrix, as shown in Figure~\ref{fig:CameraOptics}b. 
From this, we calculate the angle $\theta$ between the incoming ray and the optical axis:
\begin{equation}\label{eq:theta}                
\tan \theta=\frac{\sqrt{X^2+Y^2}\tan \theta_{\rm max}}{\sqrt{(N_{X}/2)^2+(N_Y/2)^2}}\,.
\end{equation}
Thus, the distance of the luminous point from the optical axis is
\begin{equation}\label{eq:RadialDistAir}
    \sqrt{x^2+y^2}=|z|\tan{\theta}\,.
\end{equation}
If the visualised object is submerged in a liquid of refractive index $n_\mathrm{f}$, whose surface is at a distance $h_\mathrm{c}$ below the effective pinhole of the camera (Figure~\ref{fig:CameraOptics}c), equation \ref{eq:RadialDistAir} becomes
\begin{equation}\label{eq:RadialDistLiquid}
    \sqrt{x^2+y^2}=h_\mathrm{c}\tan{\theta}+(|z|-h_\mathrm{c})\tan{\theta'}\,,
\end{equation}
where, from Snell's law
\begin{equation} \label{eq:Snell}
    \theta'=\arcsin{\left(\frac{\sin{\theta}}{n_\mathrm{f}}\right)}.
\end{equation}

\subsection{Optics of projector-object system}
\label{s:projector}

\begin{figure}[t!]
    \centering
    \includegraphics[width=\linewidth]{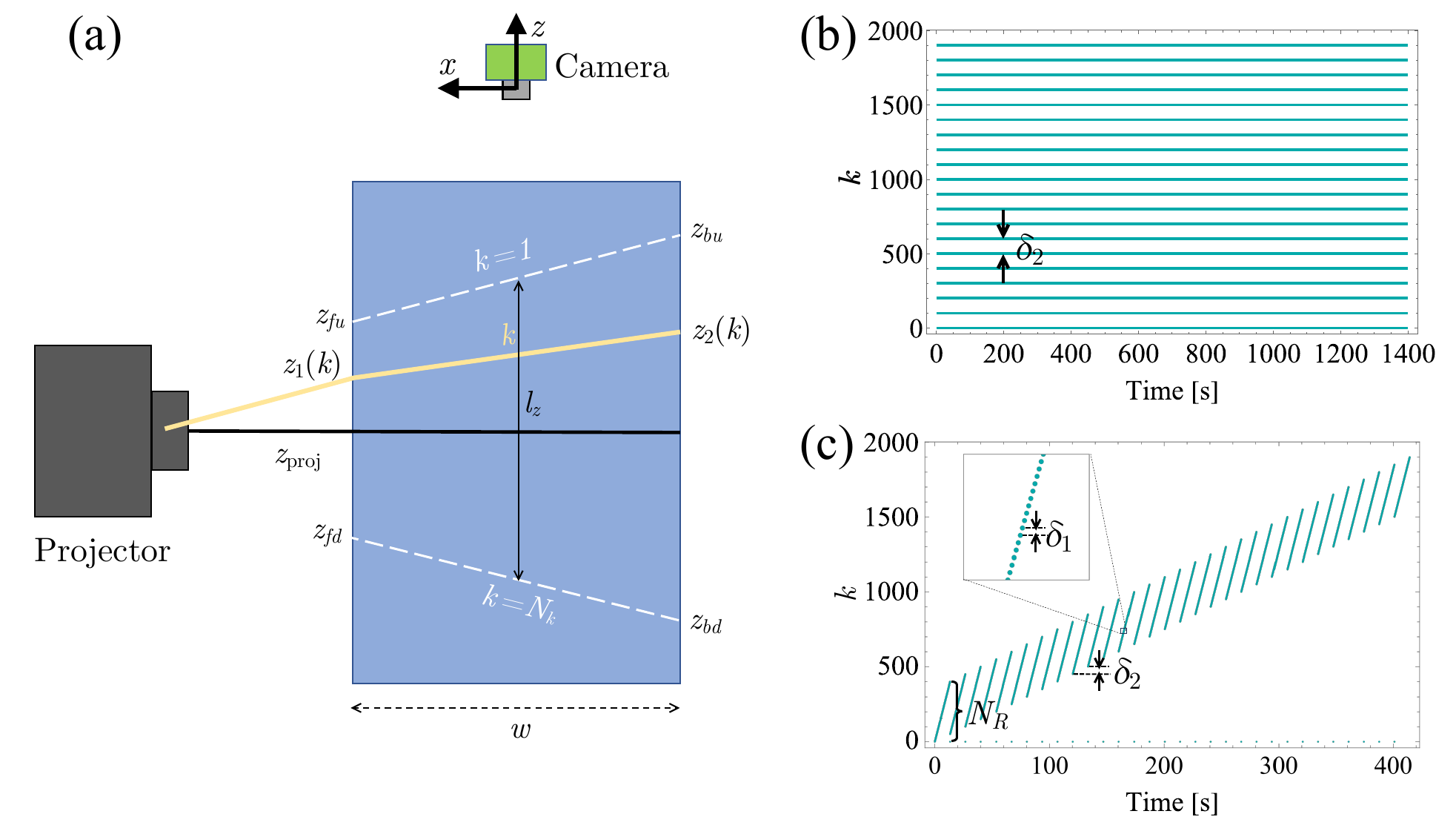}
    \caption{(a) Schematic diagram of the projector system in side view. (b,c) Spatio-temporal diagram of the illumination cast by the HD projector, where $k(t)$ is the coordinate of the illuminated rows of pixels; (b) operation in static mode where multiple rows of pixels spaced $\delta_2$ apart are displayed simultaneously; (c) operation in dynamic mode where rows of pixels separated by $\delta_1$  are sequentially illuminated to performs a scan in the region of interest of size $N_\mathrm{R}$. Upon completion of each scan, the projector flashes and the region of interest is shifted up by $\delta_2$ pixels}
    \label{fig:Projector}
\end{figure}

So far we have constrained the position of a luminous point to a ray of light received by the camera, but to close the system of equations (\ref{eq:barrel})-(\ref{eq:Snell}), we need to determine the $z$-coordinate of the luminous point.  This requires 
an expression for $z(x,y,k)$ describing the light sheet cast by the $k^{\rm th}$ row of pixels of the projector array. 
The cuboidal region of interest (ROI) in which the object is visualised (indicated by the blue rectangle in Figure~\ref{fig:Projector}a) corresponds  in our setup to the interior of the tank where the elastic disk is submerged in liquid. We align the projector horizontally so that the middle light sheet enters and exits this region at the same vertical level.  
It is also important to position and orient the projector such that it does not produce a trapezoidal image in the $(y,z)$ plane ($y$ is out of page in Figure~~\ref{fig:Projector}a).
We minimise the trapezoidal distortion so that the rows of pixels are not tilted in the $(y,z)$ plane, and thus, the planes of light are inclined only in the $(x,z)$ plane, which is the plane of Figure~\ref{fig:Projector}a.
The portrait image consists of $N_{k}$ rows of pixels. The first and last rows of pixels cast sheets of light, which enter the region at $z_\mathrm{fu}$ and $z_\mathrm{fd}$ at the front and exit at the back of the region at $z_\mathrm{bu}$ and $z_\mathrm{bd}$, respectively. \textcolor{NewText}{The values $z_\mathrm{fu}$, $z_\mathrm{fd}$, $z_\mathrm{bu}$ and $z_\mathrm{bd}$ were measured with a ruler immersed in the fluid, thus bypassing the need for solving for the additional refractions introduced by the presence of thick walls of the tank.}
The equations describing the entry point $z_1(k)$ and exit point $z_2(k)$ of the $k^{\rm th}$ row of pixels are
\begin{equation}
    z_1(k)=z_\mathrm{proj}+\left(1-\frac{2k}{N_k+1}\right)\left(\frac{z_\mathrm{fu}-z_\mathrm{fd}}{2}\right) ,
\end{equation}
\begin{equation}
    z_2(k)=z_\mathrm{proj}+\left(1-\frac{2k}{N_k+1}\right)\left(\frac{z_\mathrm{bu}-z_\mathrm{bd}}{2}\right),
\end{equation}
which we use to obtain the equation for the sheet of light associated with the $k^{\rm th}$ row,
\begin{equation}\label{eq:z}
z(x,y,k)=\frac{z_1(k)+z_2(k)}{2}+x\frac{z_1(k)-z_2(k)}{w},
\end{equation}
where $w$ indicate the width of the ROI along the $x$ direction. 
Solving equations (\ref{eq:barrel})-(\ref{eq:z}) gives a unique point $(x,y,z)$ associated with a pixel $(\xi,\eta)$ when that point is illuminated by projector's $k^{\rm th}$ row of pixels. We refer to the solution of these equations as the reconstructor function. \textcolor{NewText}{This reconstruction assumes that refraction of light on the object is negligible, which is the case when the object is index-matched with the medium or when the sheet is sufficiently thin. An upper-bound estimate of the displacement of light on a sheet of uniform thickness $b$ is set by $b \sin \psi$ in the limit of infinite refractive index, where $\psi$ is the angle of incidence. Our elastic disk has $b=50\,\mu\mathrm{m}$, which means that even if the sheet was folded so that a ray traversed it four times, the displacement of light due to refraction would be at most $200\,\mu\mathrm{m}$. This is less than the thickness of a light sheet and thus unlikely to be significant. However, when the object is not index-matched with the fluid and its surface is smooth (i.e. produces significant glare), reflections at incidence angles close to $\psi\approx90^\circ$ will deflect the beam by $2(90^\circ-\psi)$ which may cause shadows on certain regions of the object or unwanted spurious illuminations of its regions lying out-of-plane of the light sheet.}

\subsection{Scanning the object}
\label{s:ProjectorOpticsPolicy}

\begin{figure} 
    \centering
    \includegraphics[width=0.7\textwidth]{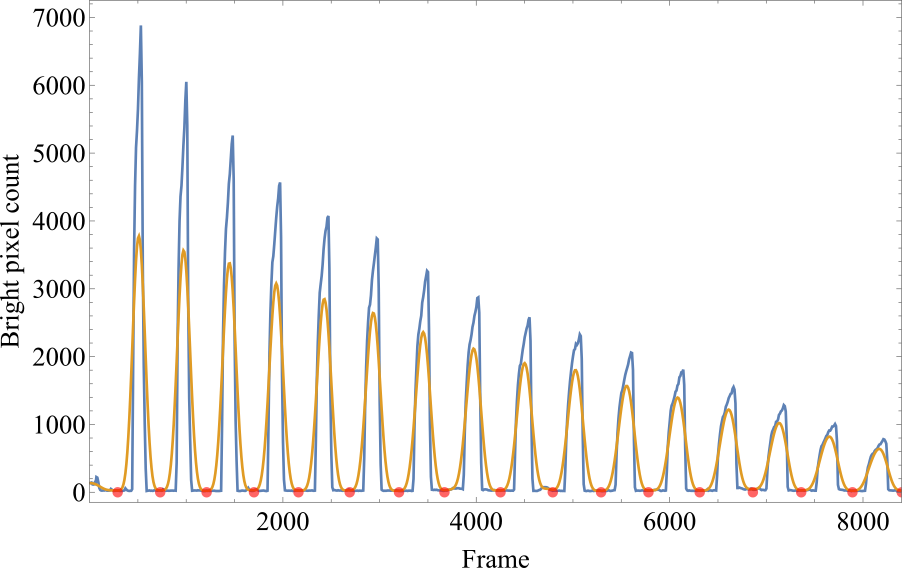}
    \caption{The time series of the total brightness of a binarized image as the object passes through stationary planes of light (blue line). The orange line shows the low-pass filtered time-series, whose minima (red points) are used to split the video into separate scans in the static mode.}
    \label{fig:intervals}
\end{figure}

We use a stack of light sheets cast by a subset of the $N_k$
horizontal rows of the projector's rectangular pixel array to obtain three-dimensional scans of the sedimenting
elastic disk. The scans can be acquired in two different modes.

In the static mode, the object moves through a fixed number
of stationary light sheets which remain illuminated, as shown in
Figure~\ref{fig:Projector}b. The light sheets emanate from horizontal rows of pixels, $\delta_2$ pixels apart, where $\delta_2$ is chosen such that the minimal separation between the light sheets inside the ROI is greater than the maximum dimension of the object. Each light sheet therefore generates one scan, completed when the object has completely traversed it. Since the sheets are illuminated permanently, the spatial resolution of the scan is determined by the frame rate of the camera; a higher frame rate is required to maintain the same spatial resolution for bodies that sediment more
quickly.

Since the camera is oriented at an approximately right
angle to the light sheets, it does not show directly when the
object has completed its motion through a given light sheet. However, given that our visualisation relies on the light
scattering from the object, the traversal of the object through the light sheets is associated with significant variations in the overall brightness of the recorded images -- highest when a large part of the object is intersected by a light sheet; lowest when the object is between two adjacent light sheets. This is illustrated in Figure~\ref{fig:intervals} where the blue line shows the brightness of the image in terms of the number of bright pixels 
for each frame in the video. The orange line was obtained by using a low-pass filter with a cut-off frequency 30 times smaller than the frame rate. This curve has clearly defined minima (highlighted by the red symbols) which identify instances when the object is located between light sheets. We use these minima to divide the frames in the video into groups associated with separate scans. This allows us to establish the z-coordinates of the light sheet that generated each scan.

We note that in the static mode, slowly moving objects take a long time to scan, and, as a result, details of their deformation and/or reorientation may be missed if they occur over timescales that are much shorter than the time required for the object to traverse a given light sheet. While the method is easy to set up, it is therefore best suited for cases where the moving object rotates/deforms relatively slowly. 

In the dynamic mode, the projector sequentially
illuminates single horizontal rows of its pixel array, separated
vertically by $\delta_1$ pixels, with the frame rate of the camera $f$ synchronised with the time interval between the illumination of
subsequent light sheets. 

In this mode, a scan is complete when one sequence of consecutively illuminated light sheets has been completed; thus the frames associated with each scan are known without any need for post-processing the video. The spatial resolution of a scan is controlled by $\delta_1$, with the resolution being higher for smaller $\delta_1$.

The method can be optimised by activating only those light sheets that cover the object in its current location. This is illustrated in Figure~\ref{fig:Projector}c for the case where the
object moves with an approximately constant velocity: following the completion of one scan (using light sheets emanating from $N_R \ll N_\zeta$ horizontal pixel rows, separated from each other by $\delta_1$ rows), the start of the next scan is initiated with an offset of $\delta_2$ pixel rows, with $\delta_2$ chosen so that the moving object remains covered by the $N_R$ light sheets. This approach requires an estimate of the
object's average speed (e.g., from preliminary experiments) and a certain margin to ensure that variations in the object's speed do not cause it to drift outside the volume illuminated by the active light sheets. For highly variable speeds, the centre of the object would have to be tracked, but for our applications this was not found to be necessary. 

The dynamic mode allows rapid scanning of slowly moving (or even
stationary) objects and is therefore preferable for cases 
where the moving object rotates/deforms relatively quickly \textcolor{NewText}{compared to the timescale of its vertical translation. To resolve such rotation/deformation, the time between scans, $\mathcal{T}=N_R/(\delta_1 f)$, must be sufficiently short compared to the timescales of the rotation/deformation; otherwise our reconstruction algorithm discussed in \S\ref{s:Reconstruction} may interpolate the shape of the object inaccurately (see Appendix \hyperref[AppendixC]{C}). In our experiments, the timescales of the rotation and deformation were estimated from preliminary observations of the sedimentation process}. 

\subsection{Image processing, data extraction and scaling}
\label{s:ImageProcessing}

\begin{figure}[t!]
    \centering
    \includegraphics[width=1\textwidth]{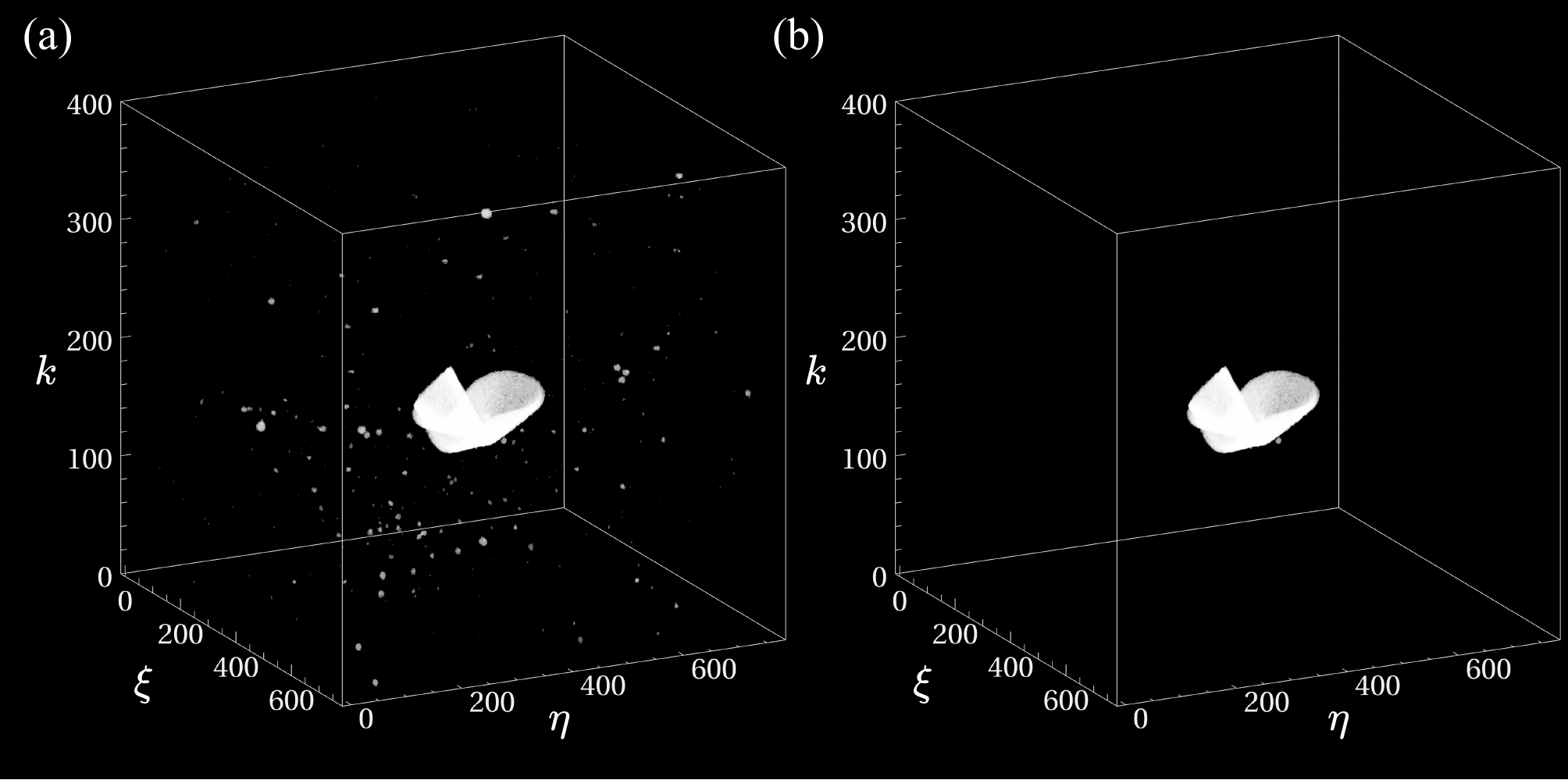}
    \caption{(a) A typical stacked image of an elastic disk, scanned in dynamic mode, following Gaussian blur convolution and prior to removal of specks of dust. (b) The filtered dataset after removal of specks of dust.}
    \label{fig:DustCleanup}
\end{figure}

Once the sedimenting disk has been scanned, the images captured by the camera require processing to mitigate bias and noise associated with the camera sensor, background illumination of the object, limited projector contrast, and the inevitable presence of illuminated specks of dust in the liquid-filled tank. 
Having calibrated the camera sensor using flat-field correction, we proceed to rescale the frames in each scan to reduce the influence of unavoidable specks of dust, which can be highly reflective, and thus, may appear brighter than the illuminated object itself. We then remove the background illumination of the object due to the projector and the environment by subtracting from each frame the average image of the scan to which this frame belongs. 
We stack the frames into a 3D image, apply Gaussian smoothing 
and binarize it to yield an image such as the one shown in Figure~\ref{fig:DustCleanup}a, where the bright regions correspond to statistically-significant signal from the elastic disk and visible dust.

To remove the specks of dust we rely on the fact that they are much smaller than the size of the object. We use two box convolutions of the image shown in Figure~\ref{fig:DustCleanup}a with different sizes of the box kernel to count the number of bright voxels within each box and threshold the resulting images to yield binary filter images, where bright specks smaller than the kernel size have been erased. A kernel size much larger than the specks of dust removes the isolated bright regions shown in Figure~\ref{fig:DustCleanup}a, while a smaller kernel size, only marginally larger than the specks of dust, is needed to eliminate those very close to the surface of the object. The filtered image shown in Figure~\ref{fig:DustCleanup}b is obtained by taking the product of the original binary image with the two filter images.  

The data is now ready for reconstruction. We determine the $(x,y,z)$ coordinates of each bright voxel in the image shown in Figure~\ref{fig:DustCleanup}b using the reconstructor function obtained in \S \ref{s:CameraOptics} and \S \ref{s:projector}. We include the time at which each voxel was scanned and stack all the scans into a four-dimensional cloud of $N_{\rm hyper}$ points $(x_{ij},y_{ij},z_{ij},t_{ij})$ where $i$ labels individual scans and $j$ labels individual points in the $i^{\rm th}$ scan. We refer to this dataset as the {\it hypercloud}.

In the next section we will develop a machine-learning algorithm to
obtain a continuous representation of the disk's motion and
deformation from the discrete data contained in the hypercloud.
The accuracy and speed of convergence of such algorithms
can be significantly reduced if the inputs are highly correlated
or unscaled \cite{LeCun2012}. When sedimenting, the elastic disk
translates by a considerable distance in the $z$-direction,
leading to a strong correlation between the input's $z$ and $t$
components. Prior to the application of the machine-learning algorithms we
therefore remove the translational motion from the dataset and also
rescale the coordinates by the size of an approximate bounding
box that contains the sedimenting disk throughout its motion.
For this purpose we approximate the motion of the disk's ``centre''
by fitting low-order polynomials,
$\overline{x}(t), \overline{y}(t)$ and $\overline{z}(t)$
to the average spatial position of the points in the hypercloud. So
if $\overline{x}(t) = C_{x0} + C_{x1} t + C_{x2} t^2 + ...$ we determine
the coefficients $C_{x0}, C_{x1}, C_{x2}, ...$ such that the quantity
\be
\frac{1}{2} \sum_{i,j} \big(x_{ij} - \overline{x}(t_{ij}) \big)^2
\ee
is minimised. \textcolor{NewText}{We repeat the procedure for $\overline{y}(t)$ and $\overline{z}(t)$}. The use of 5-th order polynomials was
found to be sufficient for all the cases we considered.
We then determined the size $\overline{b}$ of an approximate
cubic bounding box that contains the disk as it translates through space as
\be
\overline{b} = 2 \sqrt{\frac{1}{N_{\rm hyper}} \sum_{i,j} \left[
    \big(x_{ij} - \overline{x}(t_{ij}) \big)^2 +
    \big(y_{ij} - \overline{y}(t_{ij}) \big)^2 +
    \big(z_{ij} - \overline{z}(t_{ij}) \big)^2
    \right].
}
\ee
This allows us to normalise the hypercloud by scaling the
coordinates and time as
\be
\label{scale_hypercloud}
\widetilde{x}_{ij} = \frac{x_{ij} -
  \overline{x}(t_{ij})}{\overline{b}}
\mbox{ \ \ \ and \ \ \ }
\widetilde{t}_{ij} = 1 + \frac{t_{ij}}{2{\cal T}},
\ee
where ${\cal T}$ is the duration of a single scan. In the subsequent 
sections we will deal exclusively with this scaled data, and it is
understood that the actual shape and position of the disk can be
recovered by the trivial inversion of the transformations in
(\ref{scale_hypercloud}). 

\section{Surface reconstruction} 
\label{s:Reconstruction}
\subsection{The auto-encoder}
Given the scaled hypercloud containing $N_{\rm hyper}$ discrete points
$(\widetilde{x}_{ij},\widetilde{y}_{ij},\widetilde{z}_{ij},\widetilde{t}_{ij})$,
with $i$ labeling the scan and $j$ labeling the point in the scan,
the aim of the shape reconstruction is to determine a continuous vector-valued
function $\widehat{\widetilde{\bf r}}(\widetilde{u},\widetilde{v},\widetilde{t})$
to a point on the disk at an arbitrary value of the continuous
scaled time $\widetilde{t}$. Here $\widetilde{u}$ and $\widetilde{v}$
are two continuous surface coordinates that parametrise the disk. The choice of these
coordinates is, of course, not unique: even if $\widetilde{u}$ and
$\widetilde{v}$ were taken to be Lagrangian (body-fitted, material)
coordinates, we could perform the parametrisation in terms of a
Cartesian, a plane polar, or any other non-degenerate two-dimensional
coordinate system. 

We determine the function
$\widehat{\widetilde{\bf r}}(\widetilde{u},\widetilde{v}, \widetilde{t})$
by using a so-called auto-encoder which links two functions,
an encoder ${\mathbb F}_{\rm encoder}$
and a decoder ${\mathbb F}_{\rm decoder}$. Given the position of
a point on the disk in terms of the
scaled coordinates, $\widetilde{\bf r} = (\widetilde{x},\widetilde{y},
\widetilde{z})$, observed at the scaled time, $\widetilde{t}$,
the role of the encoder is to map $(\widetilde{\bf r}, \widetilde{t})$
to two surface coordinates $(\widetilde{u},\widetilde{v})$. The
role of the decoder is to map
$(\widetilde{u},\widetilde{v},\widetilde{t})$
to a vector
$\widehat{\widetilde{\bf r}}(\widetilde{u},\widetilde{v},\widetilde{t})$.
\be
\boxed{
  {\mathbb F}_{\rm encoder}:
  \left(
  \begin{array}{c}
    \widetilde{x} \\
    \widetilde{y} \\
    \widetilde{z} \\
              {\color{red}\widetilde{t}}
  \end{array}
  \right)
  \longrightarrow
  \left(
  \begin{array}{c}
    \widetilde{\color{blue}u} \\
    \widetilde{\color{blue}v}
  \end{array}
  \right)
}
\ \ \ 
\boxed{
  {\mathbb F}_{\rm decoder}:
  \left(
  \begin{array}{c}
    \widetilde{\color{blue}u} \\
    \widetilde{\color{blue}v} \\
              {\color{red}\widetilde{t}}           
  \end{array}
  \right)
  \longrightarrow
  \left(
  \begin{array}{c}
    \widehat{\widetilde{x}} \\
    \widehat{\widetilde{y}} \\
    \widehat{\widetilde{z}} 
  \end{array}
  \right)
}
\ee
Let us now consider chaining the two functions together such that
a point $(\widetilde{x}, \widetilde{y}, \widetilde{z})$
on the disk at time $\widetilde{t}$ is used as the input to
${\mathbb F}_{\rm encoder}$, and its output $(\widetilde{u},\widetilde{v})$,
together with $\widetilde{t}$ is fed into $\mathbb{F}_{\rm  decoder}$. 
If the output from the decoder recovers the input, i.e.
$\widetilde{x}=\widehat{\widetilde{x}},
\widetilde{y}=\widehat{\widetilde{y}}$
and $\widetilde{z}=\widehat{\widetilde{z}}$ then $\mathbb{F}_{\rm
  decoder}$ provides the required continuous mapping
$\widehat{\widetilde{\bf r}}(\widetilde{u},\widetilde{v}, \widetilde{t})$. 

To determine ${\mathbb F}_{\rm encoder}$ and ${\mathbb F}_{\rm decoder}$ 
we approximate them by two neural networks, ${\mathbb
  N}_{\rm encoder}$ and  ${\mathbb N}_{\rm decoder}$, respectively,
described in detail below, and determine their
weights and biases by requesting that they minimise the Mean
Pointwise Euclidean Distance (MPED) between the input and output when
the two networks are chained together and operate on the data in
the scaled hypercloud. Thus we aim to minimise
\begin{equation}
  \label{eq:MPED}
  \mathrm{M}=\frac{1}{N_\mathrm{hyper}}\sum_{i,j} \sqrt{
    \left(\widetilde{x}_{ij}-\widehat{\widetilde{x}}_{ij}\right)^2+
    \left(\widetilde{y}_{ij}-\widehat{\widetilde{y}}_{ij}\right)^2+
    \left(\widetilde{z}_{ij}-\widehat{\widetilde{z}}_{ij}\right)^2},
\end{equation}
where $\left(\widehat{\widetilde{x}}_{ij}, \widehat{\widetilde{y}}_{ij},
\widehat{\widetilde{z}}_{ij}\right)$ is the chained output from ${\mathbb
  N}_{\rm decoder}$ when the point
$(\widetilde{x}_{ij}, \widetilde{y}_{ij}, \widetilde{z}_{ij},
\widetilde{t}_{ij})$ from the scaled hypercloud is used as
the input to ${\mathbb N}_{\rm encoder}$.

Before discussing details of the network architecture and the training
process, we have to address an issue arising from the ambiguous nature of the
surface coordinates. The fact that these coordinates are in general
not Lagrangian (implying that a fixed pair of surface coordinates,
$(\widetilde{u},\widetilde{v})$, does not remain associated with a fixed
material point in the deforming and translating disk), we also have to
determine the region of the  $(\widetilde{u},\widetilde{v})$
parameter space that actually represents the disk at a given
moment in time. While a well-trained network (characterised by a small
value of $M$) ensures that the surface described by
$\widehat{\widetilde{\bf r}}(\widetilde{u},\widetilde{v}, \widetilde{t})$
will be close to the points in the hypercloud that were recorded
at time $\widetilde{t}$, the function $\widehat{\widetilde{\bf r}}$
can be evaluated for arbitrary values of $\widetilde{u},\widetilde{v}$
and thus reach points that are far outside the actual disk.

We will fully address this problem in \S\ref{s:Boundaries} below
but first introduce two methods to ensure that points in the scaled
hypercloud are at least mapped to a finite range of
$(\widetilde{u},\widetilde{v})$ coordinates:

\begin{description}
\item[Method 1] One option is to ensure that the output from the encoder is
designed such that $\widetilde{u}$ and $\widetilde{v}$ remain
bounded. We achieved this by using a $\tanh$ activation function
in the final hidden layer of the encoder. This ensures that all
points in the scaled hypercloud get mapped to surface coordinates
in the range $\widetilde{u}, \widetilde{v} \in [-1,1]$.

\item[Method 2] An alternative is to leave the output from the encoder unbounded but
to add suitable penalties to the cost function in order to bias the training
process so that the surface coordinates reflect certain known characteristics
of the disk's deformation. For instance, in our application we know
that in its undeformed configuration the disk is a planar circular
disk with radius $R$, implying that (apart from rigid body
displacements and rotations) its scaled shape can be described as
\be
\widetilde{\bf r}_0(\widetilde{u},\widetilde{v}) =
\widetilde{u} \ {\bf e}_x + \widetilde{v} \ {\bf e}_y,
\ee
where $\widetilde{u}^2 + \widetilde{v}^2 \le (R/\overline{b})^2$,
the factor $\overline{b}$ coming from the scaling of the
hypercloud via equation (\ref{scale_hypercloud}).
This establishes $\widetilde{u}$ and $\widetilde{v}$ as Cartesian
coordinates.

We now exploit that thin-walled elastic structures have an
extensional (membrane) stiffness that greatly exceeds their
bending stiffness, implying that they deform approximately
isometrically, so that material lines on the disk undergo little stretching
and shearing. We can therefore force $\widetilde{u}$ and
$\widetilde{v}$ to behave as Lagrangian coordinates by insisting
that, on the deformed disk, $\widetilde{u}$ and $\widetilde{v}$ continue
to act as arclengths, and that the $\widetilde{u}$ and $\widetilde{v}$
coordinate lines remain orthogonal. This can be enforced by
introducing the penalty
\begin{equation}
  \label{eq:penalty1}
  \mathrm{P}_1=\frac{1}{N_\mathrm{hyper}}\sum_{i,j}
  \left(\arcsin^2\left(
  \frac{(\boldsymbol{a}_u)_{ij}.(\boldsymbol{a}_v)_{ij}}
       {|(\boldsymbol{a}_u)_{ij}||(\boldsymbol{a}_v)_{ij}|}\right)
       +\left(|(\boldsymbol{a}_u)_{ij}|-1 \right)^2+
       \left(|(\boldsymbol{a}_v)_{ij}|-1 \right)^2\right)^{1/2},
\end{equation}
where $\boldsymbol{a}_{u} = \partial \widetilde{\bf r}/\partial \widetilde{u}$
and $\boldsymbol{a}_{v} = \partial \widetilde{\bf r}/\partial \widetilde{v}$
are tangent vectors to the deformed disk in the direction of the
$\widetilde{u}$ and $\widetilde{v}$ coordinate lines, respectively.

Having thus imposed the orthonormality of the coordinate lines, we can
ensure that the coordinates of points on the disk retain their
original range by introducing a second penalty
\be
P_2 = \frac{1}{N_\mathrm{hyper}} \sum_{i,j} \max \left( 0, \widetilde{u}_{ij}^2 +
\widetilde{v}_{ij}^2 - (R/\overline{b})^2 \right),
\ee
which vanishes when the encoder maps all points in the scaled
hypercloud into a circle in the $(\widetilde{u}, \widetilde{v}$)
coordinate space.
\end{description}

Both methods ensure that the deformed disk is parametrised by a finite
range of surface coordinates which facilitates the identification
of the disk's boundaries, described in \S
\ref{s:Boundaries} below. The second method has the additional
advantage of ensuring that the recovered shape is constrained
to be approximately isometric, thus mimicking the behaviour of the
actual disk. We will show in \S
\ref{sect:penalisation_matters} below that this greatly
facilitates the reconstruction of strongly deformed disks. 

When training the network we therefore use the cost function
\be
\label{combined_cost_function}
C = M \  \left(1 + \alpha_1 \frac{P_1}{P_1^{[0]}} + \alpha_2
\frac{P_2}{P_2^{[0]}} \right),
\ee
where $P_1^{[0]}$ and $P_2^{[0]}$ are the values of the two
penalties at the beginning of the training, and the weighting
factors $\alpha_1$ an $\alpha_2$ allow an
adjustment of their relative importance. We
found that using $\alpha_1=\alpha_2=1$ was sufficient.

We note that in (\ref{combined_cost_function}) the penalties
are applied in a multiplicative form.
This ensures that they play an important role
only during the early stages of the training process when
the deviation of the shape from the scaled hypercloud,
characterised by $M$, is large. Our numerical experiments suggest
that during this stage the presence of ${P_1}/{P_1^{[0]}}$ and
${P_2}/{P_2^{[0]}}$, both of which are initially $O(1)$,
suffices to guide the training process towards shapes with
the correct connectivity/topology. Once $M$ has become
sufficiently small, the role of the penalties becomes less
important.

\textcolor{NewText}{It is possible to add a penalty on curvature to prevent the parametrisation from producing high-frequency artefacts which may either arise at the early stages of training, or due to \textit{overfitting}, when the model learns the noise. In our experiments, however, we did not find any benefit from including such a penalty for three reasons: (a) The model we propose in \S\ref{sect:network} is robust to overfitting due to its number of free parameters being much smaller than the typical number of data points; (b) computing curvatures approximately doubles the training time; (c) for any non-flat surface, imposing such a penalty would work against the objective of minimising the distance $M$ thus forcing the optimiser to seek a compromise rather than fitting to the data.}

\subsection{\label{sect:network}Network architecture and training}
\begin{figure}
  \begin{center}
    \includegraphics[width=\textwidth]{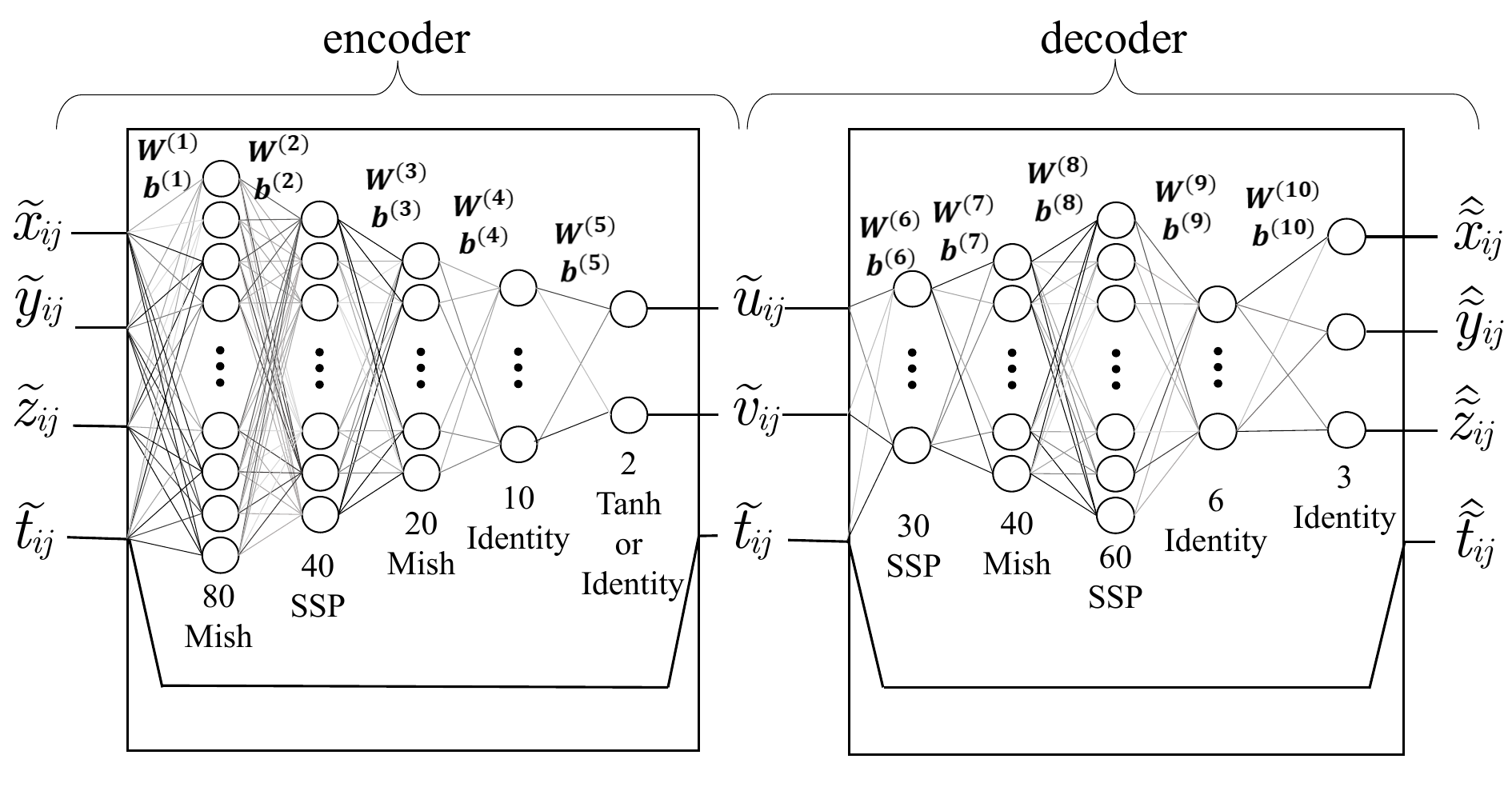}
  \end{center}
  \caption{\label{fig:neural_network} The architecture of the 
    auto-encoder which combines two neural networks, the encoder
    ${\mathbb N}_{\rm encoder}$ and the decoder ${\mathbb N}_{\rm
      decoder}$. Each layer of neurons is represented by columns of
    circles, with the number of neurons in the layer and the activation
    function indicated underneath. The encoder takes a point
    $(\widetilde{x}_{ij},\widetilde{y}_{ij},\widetilde{z}_{ij},
    \widetilde{t}_{ij})$ as its input and computes the corresponding
    surface coordinates $(\widetilde{u}_{ij},\widetilde{v}_{ij})$.
    The decoder takes the output of the encoder, augmented with
    $\widetilde{t}_{ij}$, and maps
    $(\widetilde{u}_{ij},\widetilde{v}_{ij},
    \widetilde{t}_{ij})$ to a point
    $(\widehat{\widetilde{x}}_{ij},\widehat{\widetilde{y}}_{ij},
    \widehat{\widetilde{z}}_{ij})$ on the reconstructed surface.
    The activation function in the final layer of the
    encoder is the $\tanh$ function if the penalty terms $P_1$ and
    $P_2$ are omitted (by setting $\alpha_1=\alpha_2 =0$ in the
    definition of the cost function (\ref{combined_cost_function})).
    Otherwise the identity function is used. 
    }
\end{figure}
Figure \ref{fig:neural_network} shows a sketch of the neural networks
employed in our auto-encoder. Given a point
$(\widetilde{x}_{ij},\widetilde{y}_{ij},\widetilde{z}_{ij},\widetilde{t}_{ij})$,
from the scaled hypercloud, the encoder feeds this input through
five internal layers of neurons. Each layer takes the output from the previous
layer, ${\bf X}^{(n-1)}$, and passes its output
\begin{equation}\label{eq:Xn}
  \boldsymbol{X}^{(n)}=f^{(n)}\big(\boldsymbol{W}^{(n)}
  \boldsymbol{X}^{(n-1)}+\boldsymbol{b}^{(n)}\big)\,\,,
\end{equation}
to the next layer. Here $\boldsymbol{W}^{(n)}$ is a tunable weight
matrix, $\boldsymbol{b}^{(n)}$ a tunable bias vector and $f^{(n)}$ an
activation function which acts componentwise on its vector-valued
argument. The last layer of the encoder outputs the two surface 
coordinates $(\widetilde{u}_{ij},\widetilde{v}_{ij})$.
These are recombined with $\widetilde{t}_{ij}$ and fed into the decoder
whose five internal layers compute
$(\widehat{\widetilde{x}}_{ij},\widehat{\widetilde{y}}_{ij},
\widehat{\widetilde{z}}_{ij})$. This process is applied for each
of the $N_{\rm hyper}$ points in the hypercloud to compute the cost
function (\ref{combined_cost_function}). 

In Figure \ref{fig:neural_network} the number of neurons and the activation function used is shown underneath each layer. \textcolor{NewText}{While designing neural networks is not an exact science, we used several guiding principles in our design choices. First, to avoid curvature artefacts, we limited our choices of $f^{(n)}$ only to smooth functions such as commonly used ${\rm Mish}(x)=x \tanh ({\rm SoftPlus}(x)))$, where ${\rm SoftPlus}(x)= \ln (1+e^x)$ is another commonly-used function which smoothly transitions from a zero slope in the limit of $x\to-\infty$ to a slope of 1 as $x\to + \infty $. Our second guiding principle was to use only the functions that pass through the origin ($f^{(n)}(0)=0$) and, therefore, do not introduce any large systematic bias to their output. To this end, we introduce a `Shifted' SoftPlus (SSP) defined as ${\rm SSP}(x)={\rm SoftPlus}(x)-{\rm SoftPlus}(0)$.} As discussed
above, the final layer of the encoder employs an identity activation function
if the penalties $P_1$ and $P_2$ are used to constrain the range of
$\widetilde{u}$ and $\widetilde{v}$. Otherwise the range of the
surface coordinates is constrained to the range
$\widetilde{u},\widetilde{v} \in [-1,1]$ by using a tanh activation function.

We note the presence of several layers which use identity activation functions.
These layers have no direct effect on the behaviour of the network and could in principle be eliminated but they were found to be beneficial for the training process; see \textcolor{NewText}{reference} \cite{Overparametrization2018} \textcolor{NewText}{for the documented benefits of such overparametrisation}.

The training was performed by optimising the approximately 9,000
entries in the weight matrices and bias vectors to minimise the cost function
(\ref{combined_cost_function}). This was done using an ADAM optimizer
\cite{ADAM} 
which, unlike the classical stochastic gradient descent
method, employs different and adaptively updated learning rates
for each of the network parameters. The
gradients of the cost function required by the optimiser were computed
by standard back-propagation. Prior to the start of the training
process we set all biases to zero and used Kaiming's method
\cite{Kaiming2015} to initialise the weights in the encoder; the weights
in the decoder were set using the identity initialisation,
\cite{IdentityInterpretability2021} \textcolor{NewText}{i.e., the diagonal elements in each weight matrix are 1 and off-diagonal elements are normally-distributed Gaussian noise with zero mean and 0.01 standard deviation. Together with our choice of activation functions that pass through the origin, this initialisation ensures that the decoder initially parametrises an almost flat manifold $(\widehat{\widetilde{x}},\widehat{\widetilde{y}},    \widehat{\widetilde{z}})\approx 0.25(\widetilde{u},\widetilde{v},\widetilde{t})$, which minimises the chances of the optimiser stalling at an unphysical local minimum. The pre-factor of 0.25 arises from the SSP function, which is applied twice in the decoder and has a gradient value of 0.5 at the origin, while all other activation functions have a gradient of 1.}

We performed the training in mini-batches of size  $\mathcal{B}=512$ 
and adjusted the learning rate in three stages:
1 million iterations (with updates performed after each batch)
with a learning rate $10^{-3}$, followed by 0.9 million iterations
with a learning rate of of $10^{-4}$ and
0.1 million iterations with a learning rate of $10^{-5}$.
The remaining parameters in the ADAM optimiser were kept at their
recommended default values ($\beta_1=0.9, \beta_2=0.999$
and $\epsilon=10^{-7}$; see \cite{ADAM} for details).

The data processing and training of the neural network was implemented
in Wolfram Mathematica 14.3 which contains a robust
and accessible machine-learning framework supporting GPU
acceleration. The code is provided via the GitHub repository at 
\cite{github}. 
For a typical dataset containing $N_\mathrm{hyper}=7\times 10^6$ data
points the training took approximately 40 minutes ($\sim 830$ iterations per
second) on a computer with AMD Ryzen 9 7900X CPU overclocked to
5.7~GHz, DDR5 RAM at the clock speed of 6000~MT/s and an NVIDIA GeForce
RTX3090 graphics card. The VRAM memory required by the training
process is 1.3 gigabytes, which means the training can be performed on
virtually any graphics card available on the market. The inclusion of
the penalty term  $P_1$ requires the evaluation of derivatives which
were computed using central finite differences. This slowed down the
training process by a factor of two.

\textcolor{NewText}{Following training, we extract the decoder which is the function $\widehat{\widetilde{\bf r}}(\widetilde{u}, \widetilde{v},\widetilde{t})$ and we reverse the normalisation steps described by equation (\ref{scale_hypercloud}) to create the desired function $\widehat{\bf r}(\widetilde{u}, \widetilde{v},\widetilde{t})$ which parametrises the surface in the lab-frame coordinates. Using $\widehat{\bf r}(\widetilde{u}, \widetilde{v},\widetilde{t})$ one may sample the surface to make a 3D plot of the disk 
at any specified 
$\widetilde{t}$ or compute other derived quantities such as normal vectors, principal curvatures, metric tensor etc, as desired. We have implemented the calculation of these and other quantities and provide the relevant code in the GitHub repository \cite{github}. We note that, since the hypercloud does not contain information about displacement of material points, we cannot recover in-plane stretching of the material. For instance, we cannot determine if an object of elliptical appearance is elliptical in its undeformed state, or if it is a circle that has been stretched to an elliptical shape.}

\subsection{\label{sect:validation}Validation of the training process}
\begin{figure}
  \begin{center}
    \includegraphics[width=\textwidth]{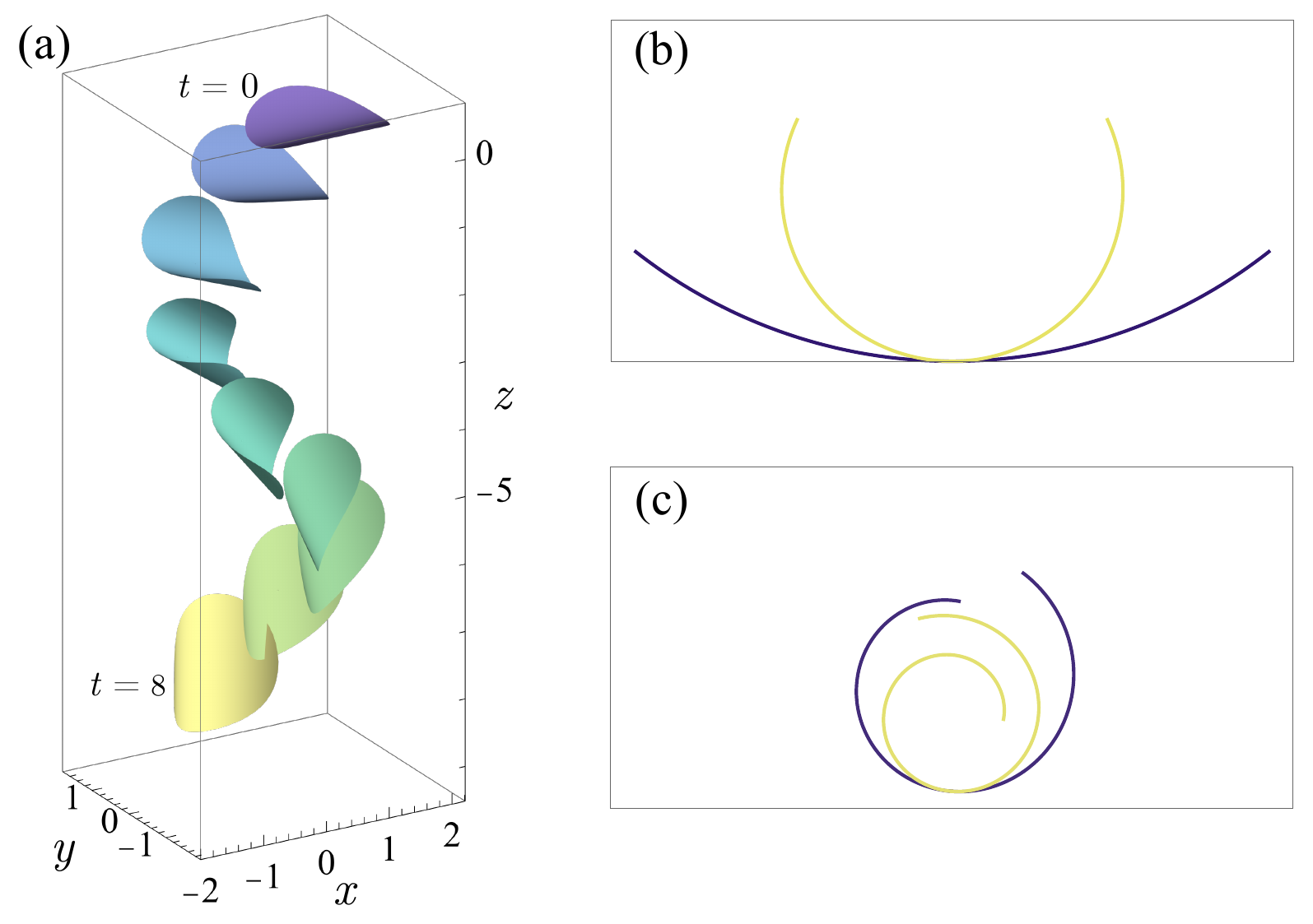}
  \end{center}
  \caption{\label{fig:synthetic_disk_plot}(a) Snapshots of the 3D disk
    shape used to analyse the robustness of the training process.
    (b,c) Snapshots of cross-sections across the isometrically
    deforming disk for the case when the reference disk is wrapped
    around (b) a cylinder of periodically varying radius $R(t)$,
    (c) a logarithmic spiral with periodically varying
    curvature $\kappa(t)$. 
 }
\end{figure}

To demonstrate the robustness of the training process we
created a synthetic hypercloud corresponding to the sedimentation of
a circular disk that translates and rotates while undergoing an isometric
deformation from its planar reference shape. This is illustrated by
the snapshots in Figure \ref{fig:synthetic_disk_plot}(a):
the disk follows a three-dimensional (but mainly downward)
trajectory, while simultaneously performing rotations about three axes. 
An isometric deformation was
imposed by wrapping the disk's reference shape around a cylinder
of periodically varying radius, $R_c(t)$, as illustrated
in the cross-sectional snapshots shown in Figure
\ref{fig:synthetic_disk_plot}(b). We also considered
the more challenging case when the disk's deformation is obtained by
wrapping it around a logarithmic spiral with periodically varying
curvature; see Figure \ref{fig:synthetic_disk_plot}(c). This allowed deformations where parts of the disk overlap.

\begin{figure}
  \begin{center}
    \includegraphics[width=0.8\textwidth]{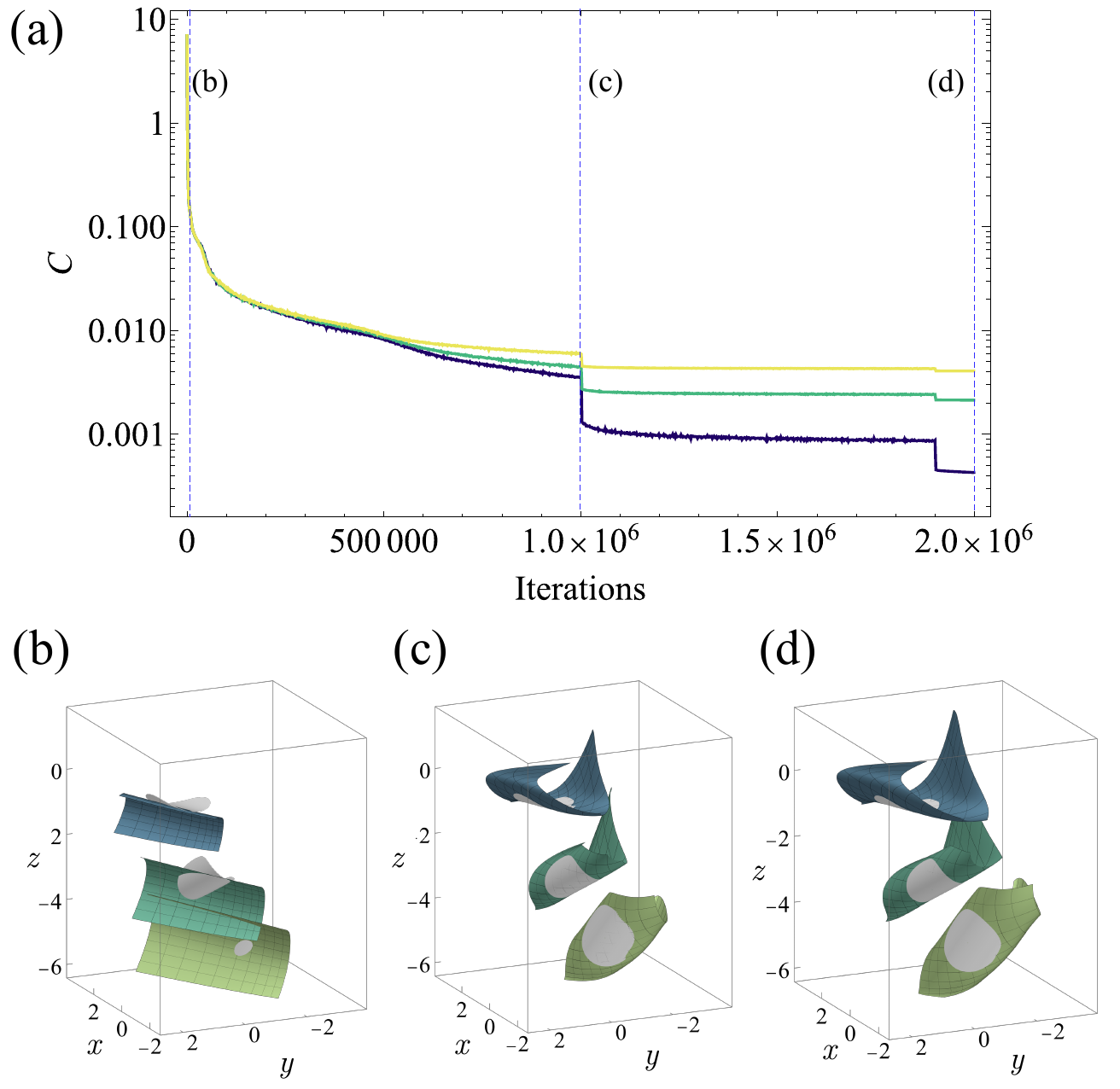}
  \end{center}
  \caption{\label{fig:training}Illustration of the training process.
    (a) Evolution of the cost $C$ versus the number of iterations
    performed by the optimiser for three different levels of noise. Dark blue: $\delta=0$; green $\delta=0.01$, yellow $\delta=0.02$.
    (b-d) Three sequences of snapshots of the disk (grey) and the predictions
    $\widehat{\widetilde{\bf r}}(\widetilde{u}, \widetilde{v}, \widetilde{t})$
    from the decoder at three different stages of the training process,
    indicated by the vertical dashed lines in (a). 
    Note that $\widehat{\widetilde{\bf r}}(\widetilde{u},
    \widetilde{v})$ is plotted over the entire range of
    the surface coordinates, $\widetilde{u},\widetilde{v} \in [-1,1]$.
    The training process proceeds so rapidly that the shapes obtained after $1.0 \times 10^6$ and  $2.0 \times 10^6$ iterations are virtually indistinguishable.}
\end{figure}

Figure~\ref{fig:training} illustrates the progress of the training
process for the deformation illustrated in
Figs. \ref{fig:synthetic_disk_plot}(a,b).
The blue line in Figure~\ref{fig:training}(a) shows
the cost $C$ as a function of the number of the iterations performed
by the optimiser for the case without penalisation, i.e. for
$\alpha_1=\alpha_2=0$ in (\ref{combined_cost_function}). Overall, the
cost function can be seen to decrease continuously, and the benefit of
reducing the learning rate after a certain number of
iterations is evident: during the first 1 million iterations,
corresponding to the stage of coarse learning, the value of $C$
initially decreases very rapidly but then begins to plateau. 
The reduction in learning rate during the second stage enables another
rapid decrease in $C$ before it approaches a second,
lower plateau. The third stage further fine-tunes the parametrisation of
the network, albeit with a significantly smaller reduction in the cost.

Figs. \ref{fig:training}(b-d) illustrate how the 
training process improves the representation of the disk by the
decoder. We show snapshots of the actual disk (grey) and the surface
$\widehat{\bf r}(\widetilde{u},\widetilde{v},
\widetilde{t})$, obtained from the decoder at three different points
in the training process, indicated by the blue dashed lines
in Figure \ref{fig:training}(a). Note that we plotted
$\widehat{\bf r}(\widetilde{u},
\widetilde{v},\widetilde{t})$ over the entire range of
the surface coordinates, $\widetilde{u},\widetilde{v} \in [-1,1]$
which, as expected, causes the reconstructed surface to extend far
beyond the actual disk. However, in the regions
where the two surfaces overlap, the shape of the disk can be seen
to be reconstructed very accurately by the fully trained network,
and there is, in fact, very little visible difference between
the shapes obtained after $1.0 \times 10^6$ and $2.0 \times 10^6$ iterations.

The other lines in Figure \ref{fig:training}(a) illustrate the
robustness of the training process with respect to noise, introduced 
by perturbing each $z_{ij}$ by a uniformly distributed random number
between $-\delta$ and $+\delta$.

The presence of noise can be seen to increase the cost $C$
which, since $\alpha_1=\alpha_2=0$, simply represents the MPED of the
decoder's predictions, $\widehat{\widetilde{\bf r}}(\widetilde{u},
\widetilde{v},\widetilde{t})$, relative to the points in
the perturbed hypercloud.
The increase therefore indicates that our network is robust to
overfitting: if the cost remained unaffected by the presence of noise,
$\widehat{\widetilde{\bf r}}(\widetilde{u}, \widetilde{v},
\widetilde{t})$ would have to closely follow the small-scale features
introduced by the presence of random spatial fluctuations in the
position of the points in the hypercloud. The fact that the cost increases
with an increase in $\delta$ implies that the neural network continues
to faithfully represent the actual underlying shape. In fact,
Figs. \ref{fig:training}(b-d) show plots
of  $\widehat{\bf r}(\widetilde{u}, \widetilde{v})$ for
the case with the largest noise; the plots of the reconstructed shapes
in the presence of noise are virtually indistinguishable from those
obtained for $\delta = 0$.

\subsection{ \label{s:Boundaries}Determination of the boundary}
The final challenge in the shape reconstruction is the
identification of the disk's boundary. We have seen in 
Figure \ref{fig:training} that, at a given time $\widetilde{t}$,
the surface described by the output from the fully-trained decoder,
$\widehat{\bf r}(\widetilde{u}, \widetilde{v}, \widetilde{t})$,
overlaps with the shape of the disk but, if
plotted over the entire range of the surface coordinates,
$\widetilde{u},\widetilde{v} \in [-1,1]$, it extends far beyond it. \textcolor{NewText}{We must therefore find the subset of the space of $(\widetilde{u}, \widetilde{v})$ which corresponds to the interior of the disk for any given time $\widetilde{t}$. Recall, however, that for any time $\widetilde{t}$, only at most a very thin (1D) slice of the disk is illuminated which is insufficient to determine the boundary solely from that single time point. Therefore, our identification of the boundary must involve some form of time-interpolation of the information obtained across multiple completed scans.}

\begin{figure}
  \begin{center}
    \includegraphics[width=\textwidth]{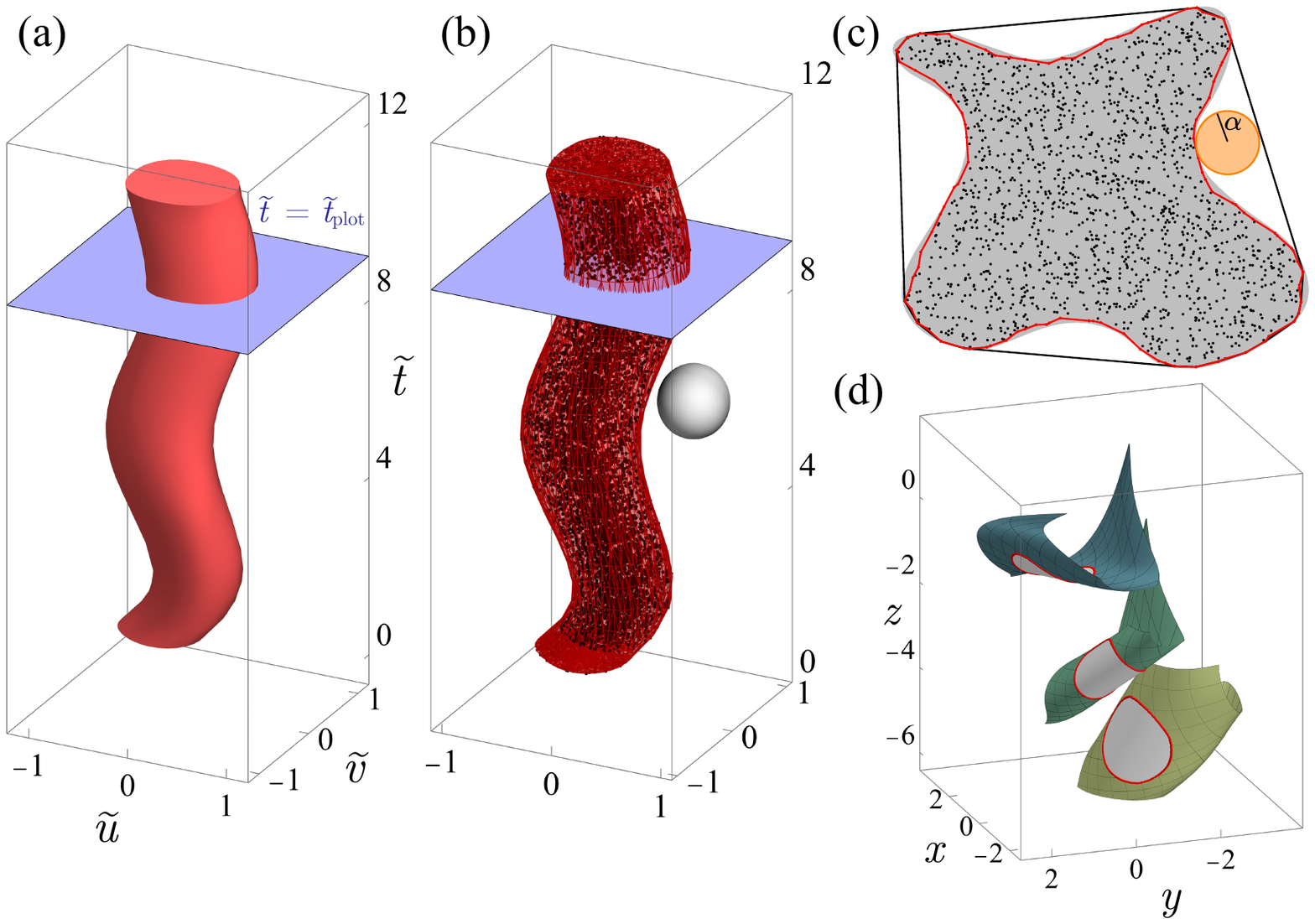}
  \end{center}
  \caption{\label{fig:boundary_determination}Illustration of the
    process used to reconstruct the disk boundary. (a) The volume $V$
    in $(\widetilde{u}, \widetilde{v}, \widetilde{t})$ space contains
    the continuous surface coordinates  $(\widetilde{u}, \widetilde{v})$ of all
    points on the disk over the duration of the experiment. At a given
    moment in time $\widetilde{t}=\widetilde{t}_{\rm plot}$, say, the
    shape of the disk can obtained by evaluating the output of
    the decoder for all values of $(\widetilde{u}, \widetilde{v})$
    contained in the intersection of $V$ (red)
    with the plane  $\widetilde{t}=\widetilde{t}_{\rm plot}$ (purple) 
    (b) The approximation of $V$ given by the $\alpha$-shape of the
    $(\widetilde{u}_{ij},\widetilde{v}_{ij}, \widetilde{t}_{ij})$
    point cloud, constructed using a ball of radius $\alpha=0.3$.
    (c) 2D sketch illustrating the construction of the
    $\alpha$-shape, surrounded by its concave hull, for a point cloud
    occupying the shaded non-convex region. (d) Snapshots of the sedimenting
    disk (as in Figure \ref{fig:synthetic_disk_plot}(a,b))
    (grey), with the reconstructed shape
    evaluated over the entire range of surface coordinates (green),
    and the boundary (red).    
  }
\end{figure}

To restrict the evaluation of $\widehat{\widetilde{\bf r}}$ to points
inside the disk, we now consider the output from the  encoder.
Given a point $(\widetilde{x},\widetilde{y},\widetilde{z})$ on the
continous disk at a continous time $\widetilde{t}$, the encoder
returns the corresponding continuous surface coordinates
$(\widetilde{u}, \widetilde{v})$. The application of the encoder to {\em all} 4-tuples
$(\widetilde{x},\widetilde{y},\widetilde{z},\widetilde{t})$,
representing the positions of all points
$(\widetilde{x},\widetilde{y},\widetilde{z})$ on the disk
at time $\widetilde{t}$ over the duration of the experiment
therefore defines a volume $V$ in a three-dimensional
$(\widetilde{u}, \widetilde{v}, \widetilde{t})$ space.
Each point in $V$ contains all the 3-tuples
$(\widetilde{u}, \widetilde{v}, \widetilde{t})$ 
that correspond to some point
$(\widetilde{x},\widetilde{y},\widetilde{z})$
on the disk at time $\widetilde{t}$.
The red volume in Figure \ref{fig:boundary_determination}(a)  shows
a plot of the region $V$ for the sedimenting disk shown in
Figure \ref{fig:synthetic_disk_plot}(a,b).
Using this volume we can recover the disk shape at a given
time $\widetilde{t} = \widetilde{t}_{\rm plot}$, say, by evaluating
the decoder for all values of $(\widetilde{u}, \widetilde{v})$
that are contained in the intersection of the volume $V$ with
the plane $\widetilde{t} = \widetilde{t}_{\rm plot}$, shown in
purple in Figure \ref{fig:boundary_determination}(a). We note that
the plot of $V$ shows clearly that the surface coordinates
$(\widetilde{u}, \widetilde{v})$ are not Lagrangian coordinates;
$\widetilde{u}$ and $\widetilde{v}$ parametrise the disk 
but their range\textcolor{NewText}{, at any given time is not a circular region and it} changes continuously with time. As a result, the
volume $V$ is generally non-convex. \textcolor{NewText}{This non-convexity becomes even more pronounced if the reconstructed object itself has a non-convex shape in its undeformed state.}

To turn these observations into a practically useful algorithm
we have to account for the fact that disk is only sampled at the
discrete points $(\widetilde{x}_{ij}, \widetilde{y}_{ij},
\widetilde{z}_{ij}, \widetilde{t}_{ij})$ contained in the hypercloud.
Applying the encoder to these 4-tuples yields the discrete points
$(\widetilde{u}_{ij}, \widetilde{v}_{ij}, \widetilde{t}_{ij})$
in the three-dimensional $(\widetilde{u}, \widetilde{v},
\widetilde{t})$ space.  We therefore have to construct an approximation
for $V$ based solely on these discrete points while
accounting for the possible non-convexity of the volume. 
We do this by constructing the so-called $\alpha$-shape
(also known as the concave hull) of the $(\widetilde{u}_{ij},
\widetilde{v}_{ij}, \widetilde{t}_{ij})$ point cloud. The idea behind
this construction is illustrated in
the conceptual two-dimensional sketch in Figure \ref{fig:boundary_determination}(c)
where the black symbols occupy a shaded non-convex region. An
attempt to reconstruct this region via a straightforward Delaunay triangulation
would embed the points in their convex hull, shown by the dark grey line.
The concave hull, shown by the red line, is obtained by ``rolling'' a
circle of radius $\alpha$ around the point cloud and connecting points
that are touched by the circle. The sketch shows that for a suitably
chosen value of $\alpha$ this method is capable of resolving the
non-convex boundary of the point cloud. The idea is easily generalised to
three dimensions (where the circle is replaced by a sphere) and
efficient algorithms for the construction of
three-dimensional $\alpha$-shapes exist. We used Mathematica's
{\tt ConcaveHullMesh} function which approximates
the potentially non-convex volume $V$ containing a three-dimensional
point cloud in terms of a collection of straight-sided tetrahedra.
Figure \ref{fig:boundary_determination}(b) shows the resulting discrete
approximation of $V$ and the sphere 
used to 
obtain the complex hull enclosing the $(\widetilde{u}_{ij},
\widetilde{v}_{ij}, \widetilde{t}_{ij})$ point cloud. \textcolor{NewText}{The computation of this concave hull is relatively quick: it takes approximately 1~minute for an experiment with 1~million data points.}

Given this approximation of the volume $V$, we can now determine the shape
of the disk at a given instant $\widetilde{t}_{\rm
  plot}$ by sampling the plane  $\widetilde{t} = \widetilde{t}_{\rm
  plot}$ at a large number of regularly spaced surface coordinates in the range
$\widetilde{u},\widetilde{v} \in [-1,1]$. For each sample point
$(\widetilde{u}_{\rm sample},\widetilde{v}_{\rm
  sample},\widetilde{t}_{\rm plot})$ we check if the
point is contained in our discrete approximation of $V$,
using a fast ``point in tetrahedron'' test. If so, we evaluate
the decoder to obtain the point on the disk from
$\widehat{\widetilde{\bf r}}(\widetilde{u}_{\rm
  sample},\widetilde{v}_{\rm sample},\widetilde{t}_{\rm plot})$.
Furthermore, we record the $(\widetilde{u},\widetilde{v})$ coordinates
of all sample points deemed to be inside $V$ and determine their
concave hull in the $(\widetilde{u},\widetilde{v})$-plane, using a
disk of the same radius $\alpha$ that we used to determine the
boundary of $V$. This concave hull provides a continous representation
of the coordinates $(\widetilde{u},\widetilde{v})$  that are mapped
to the boundary of the disk. Evaluating the decoder for these values
then identifies the boundary of the disk in 3D space. This boundary is
shown by the red lines in Figure \ref{fig:boundary_determination}(d)
for three snapshots of the reconstructed disk. The plot shows
that the method is capable of accurately determining the boundary of
the disk.

\subsection{\label{sect:penalisation_matters}The use of isometricity
  enforcing penalties for strongly deformed shapes}
\begin{figure}
  \begin{center}
    \includegraphics[width=\textwidth]{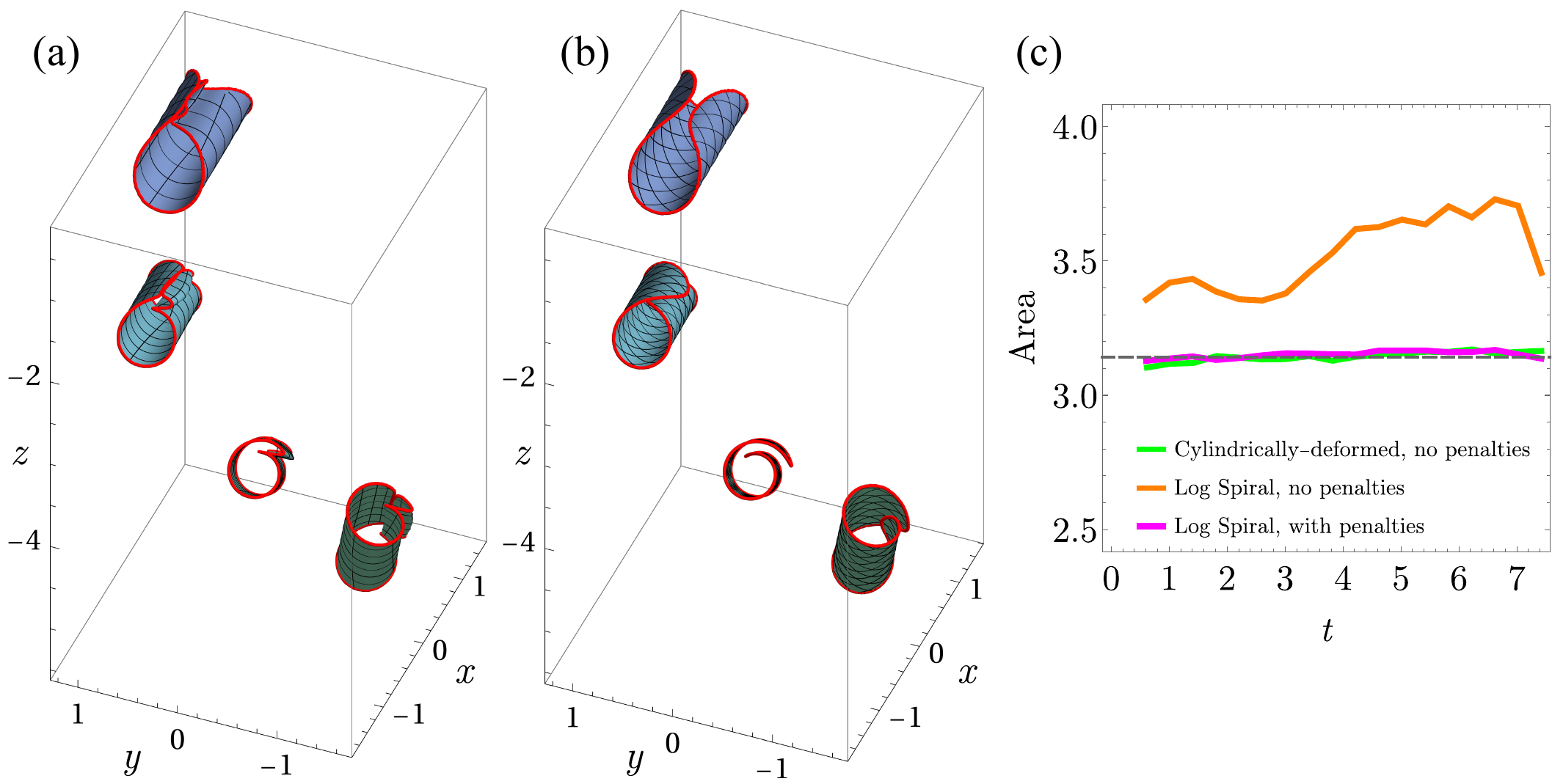}
  \end{center}
  \caption{\label{fig:log_spiral_test_case}Plots illustrating the
    efficacy of method 2 (using the penalties $P_1$ and $P_2$ to enforce the isometricity of the deformation) to deal with shapes
    where initially distant parts of the deformed disk become close to
    each other. (a,b) show the reconstruction of the synthetic data set representing a sedimenting disk undergoing the deformation shown in
    Figure \ref{fig:synthetic_disk_plot}(a,c): (a) using method 1 and and (b) method 2, where the latter incorporates the isometricity-enforcing 
    penalties $P_1$ and $P_2$ into the 
    cost function $C$.
    (c) Time trace of the area of the reconstructed disk.}
\end{figure}
In the test case considered above (and for many other tests involving relatively simple
shapes; see \cite{TymPhD}), method 1 sufficed to recover the shape of the disk. The penalties $P_1$ and $P_2$ were not required, and their inclusion into the cost function (\ref{combined_cost_function}) and the use of method 2 made no difference to the quality of the shape reconstruction.

However, the use of the penalties was found to be essential for cases
when the disk is deformed so strongly that initially distant material
points come close to each other. Figure \ref{fig:log_spiral_test_case}
illustrates the resulting failure of the reconstruction for the case when an
initially planar circular disk is wrapped into a tight logarithmic
spiral, so that its opposite ends overlap -- the case illustrated in Figure 
\ref{fig:synthetic_disk_plot}(c). Without the presence of
the isometricity enforcing penalties in the cost function
(\ref{combined_cost_function}), the training process aims to minimise
the distance of the reconstructed shape from the points
in the hypercloud. For cases, where the disk's deformation has moved
initially distant material points close to each other, 
the training process is liable to minimise the cost by introducing
spurious connections between distinct parts of the disk. Such spurious
connections can clearly be seen in  Figure \ref{fig:log_spiral_test_case}(a).

Figure \ref{fig:log_spiral_test_case}(b) shows that the inclusion of
the penalties $P_1$ and $P_2$ allows the training process to
accurately recover the
disk's actual shape without introducing such spurious connections.
To demonstrate that this improvement is indeed
due to us enforcing the isometricity
of the deformation, Figure \ref{fig:log_spiral_test_case}(c) shows a
time trace of the disk's surface area, $A(t)$, which
for the actual disk remains constant, $A(t) = \pi$.
The orange line shows that
the omission of the penalties from the cost function leads to
large variations in $A(t)$, introduced by the presence spurious connections
between different parts of the deforming disk. The imposition of
isometricity via the inclusion of the penalties results in the area
remaining very close (with mean deviation of 0.01) to the correct
value of $\pi$ (magenta line).
For reference, the plot also shows a green line which represents $A(t)$
for the deformation considered in the previous sections where
the disk was wrapped around a cylinder with periodically varying
radius. For this much simpler deformation, the reconstructed area can
be seen to be maintained at its correct constant value (with mean deviation of 0.015) even without
the imposition of the penalties.

\section{Results: Settling of an initially crumpled elastic disk}
\label{s:results}
\begin{figure}[h]
\includegraphics[width=0.7\textwidth]{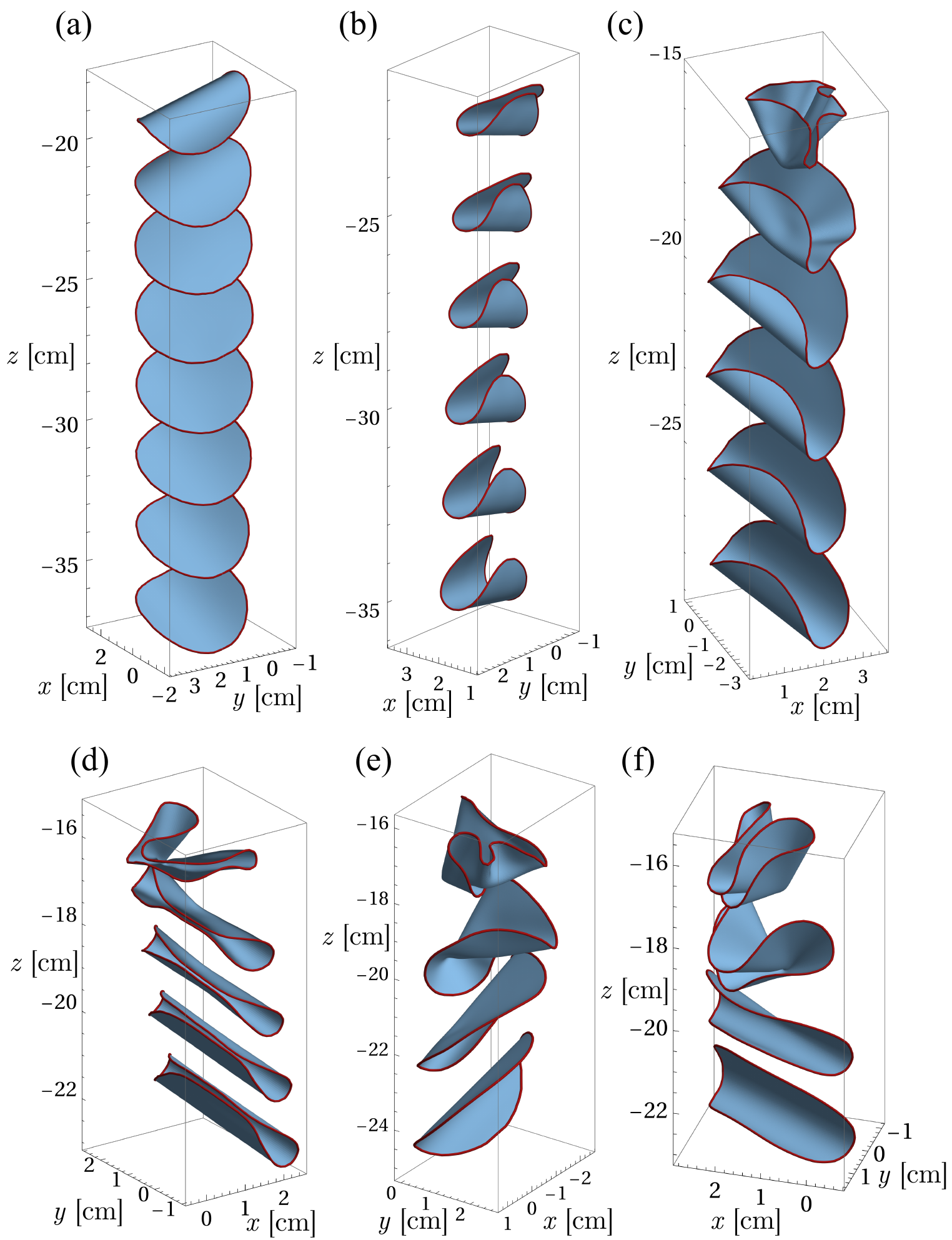}

\caption{Snapshots showing reconstructions of a deforming sedimenting disk, released from different initial configurations: (a,b) U-bent disks folded in two prior to release; (c-f) crumpled disks which were folded in four. The origin of the $z$ coordinate is at the liquid interface. Disks were typically released $15$\,cm below the surface. 
\label{fig:results}}
\end{figure}

Having validated our shape reconstruction algorithm using a synthetic dataset,
we now demonstrate the full workflow (from the data acquisition to the
shape reconstruction) in a study of sedimenting
elastic disks, performed using the experimental setup described in \S
\ref{S:Setup}: in each experiment, a single elastic disk was
immersed in the viscous fluid, folded and then released.
The disk then sedimented and reoriented itself within the fluid while
simultaneously deforming in response to gravitational and fluid
mechanical forces acting on it. The large fluid viscosity and
the small difference between the densities of the fluid and the disk
resulted in slow sedimentation in a regime where inertial forces
could be neglected. We used the theoretical prediction for the velocity of our disk falling edge on, $U= 0.281$\,mm/s \cite{Happel1983}, to estimate an upper bound for the  
Reynolds number, $Re = \rho_f U R/\mu \approx 5 \times 10^{-3}$.
The importance of fluid structure interaction is characterised by
the ratio of the disk's bending stiffness to the typical fluid
traction, represented by the parameter
\begin{equation}
B = \frac{D}{bR^3 \Delta\rho\, g} \approx 0.012,
\end{equation}
where $\displaystyle D=\frac{Eb^3}{12(1-\nu^2)}=1.2\times 10^{-8}$\,N\,m is the bending stiffness, $\Delta \rho= \rho_\mathrm{s} -\rho_\mathrm{f}$ is the excess density of the disk and $g$ is the gravitational acceleration. We recorded the disk's motion and deformation by operating the
projector in dynamic mode, with the region of interest being scanned
with $N_R = 399$ 
sequentially illuminated
light sheets, separated vertically by $\delta_1=1$ pixels. Following
the completion of each scan, the region of interest was shifted
by $\delta_2=40$ pixels; see
Figure \ref{fig:Projector}. Scans were performed at a frame rate
of $f=15$~fps, resulting in a complete scan, covering the moving
region of interest, being performed in $\mathcal{T}=26.66$~s. The region of interest
moved with an average speed of $v_\mathrm{ROI}=0.227$\,mm/s, which is close to the vertical speed of the disk, $V=0.25$\,mm/s. 

Figure~\ref{fig:results} shows examples of the reconstructed shapes,
representing snapshots of the deforming disk's motion through the tank
following its release from various initial configurations. In
Figure~\ref{fig:results}(a) we
started with a relatively easy-to-reconstruct shape where the disk is
initially bent into an upside-down U-shape. Following its release, the disk
sediments, resulting in a fluid traction that combines with the
elastic restoring forces to open up the bent disk. Once the disk reaches an
approximately flat shape, the fluid traction continues to act
upwards and thus bends the disk into the opposite direction,
ultimately resulting in the sedimentation in an upright U-shape. It is interesting
to note that during the transition through the approximately flat
state, the orientation of the disk's axis of bending changes by
approximately 90 degrees.

Figure~\ref{fig:results}(b) shows the corresponding
results for a disk that is
initially deformed into a more strongly pinched shape with its
opposite sides being in near contact. Following its release from
an orientation in which the disk is tilted sideways, the disk
reopens (indicating that the elastic restoring forces dominate the
fluid traction), while righting itself and  ultimately sedimenting in an
upright U-shape again.

\begin{figure}[tb]
    \centering
    \includegraphics[width=0.6\linewidth]{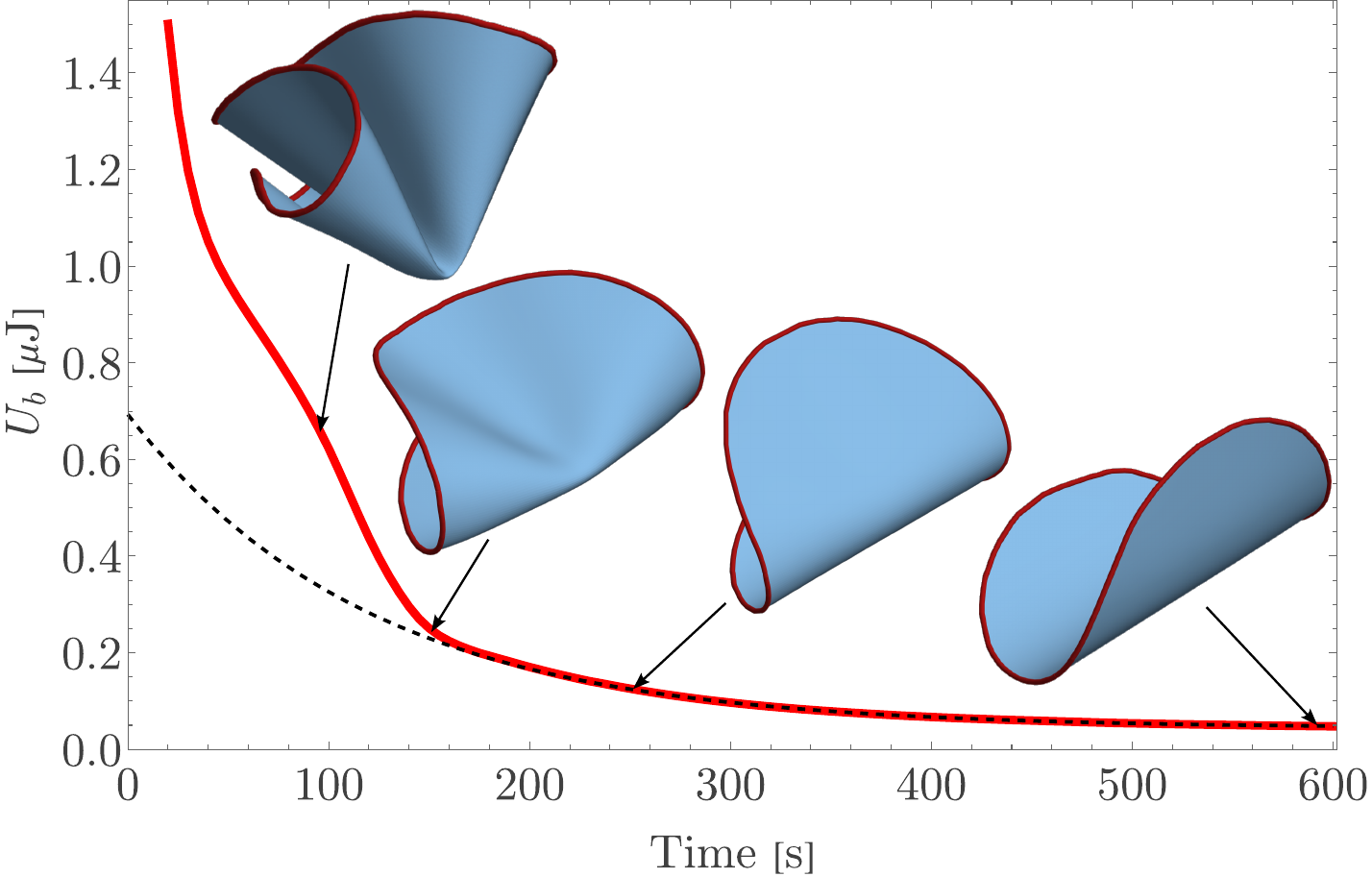}
    \caption{\textcolor{NewText}{Evolution of the bending energy $U_b(t)$ of the initially crumpled disk shown in Figure \ref{fig:results}(f). Insets show the respective reconstructions of the disk at times indicated with the arrows; the first three shapes from the left correspond to the bottom three shapes in Figure \ref{fig:results}(f) but shown from a different vantage point. 
    The dashed line shows an exponential fit to the latter stage of the experiment  ($t\geq250$~s) where the disk is no longer crumpled.}}
    \label{fig:BendingEnergy}
\end{figure}

Finally, Figures~\ref{fig:results}(c--f) 
show the most challenging initial shapes where the disk is released from a
strongly crumpled configuration. Here, and in the previous case, the
incorporation of the isometricity enforcing penalties into the cost
function is vital to avoid the occurrence of spurious connections
between opposite parts of the reconstructed disk.
\textcolor{NewText}{Using the expression for the bending energy ${U_b=\int \frac{D}{2}\left[ (\kappa _1+\kappa _2)^2-2(1-\nu )\kappa_1\kappa_2 \right] dA}$ \cite{landau1986theory}, where $\kappa_1$ and $\kappa_2$ are the principal curvatures, 
in Figure~\ref{fig:BendingEnergy} we compute the evolution of $U_b(t)$ corresponding to the experiment shown in Figure~\ref{fig:results}(f). The relaxation timescale of the bending energy from its crumpled state to a U-shape is of the order of $\sim 10^2$\,s. 
The disk shape adopts a U-shape around $t=250$\,s (see the fourth shape from the top in Figure \ref{fig:results}(f) and the third shape from the left in Figure~\ref{fig:BendingEnergy}).
From this point onward, the bending energy decays approximately exponentially as $U_b(t)\approx A e^{-t/(120\,\mathrm{s})}+0.044\,\mu\mathrm{J}$, which is indicated by the dashed line in Figure \ref{fig:BendingEnergy}. Therefore, this relaxation timescale is also on the order of $10^2$\,s. We refer to Appendix \hyperref[s:AppendixB]{B} for an a-priori estimate of this timescale.}

Given that in these examples, the shape reconstruction is based on
actual experimental data, we do not
have a gold-standard via which to assess the accuracy of the
reconstructed shapes. However, the fact that the disk is expected
to deform with little in-plane stretching implies that its area
should remain approximately constant throughout the experiment. We assessed this by determining the radius of the reconstructed disk \textcolor{NewText}{once it had uncrumpled}. We computed a value of $19.16$~mm, which differs by $<1\%$ from that of the actual disk. The reconstructed radius fluctuated across the time
slices with a standard deviation of 0.04~mm, or $0.2\%$.

\section{Discussion and Conclusions}
\label{s:conclusion}
 
We have developed a low-cost, 3D reconstruction method that uses a single camera to accurately visualise the motion and deformation of transparent, thin elastic sheets in flow. We relied on the transparency of the sheet to capture its illuminated outline due to Rayleigh scattering during scans at a rate much faster than that of the motion and deformation of the sheet. Following image processing, we obtained a spatio-temporal representation of the sheet, which we reconstructed with a neural autoencoder.

We validated our shape reconstruction algorithm using synthetic data
sets, and then demonstrated its capabilities when
operating on data obtained from actual experiments with sedimenting elastic disks. The inclusion of the isometricity-enforcing penalties into the cost function used to determine the parameters of the neural networks
enabled us to robustly reconstruct shapes, even in cases where the
disks were so strongly deformed that different parts came into
close contact with each other.

We stress that our choices for the parameter values and the
form of the various functions used in the reconstruction algorithm
were guided by much trial and error. The choices include the
method used to scale the raw data, the network layout, the choice
of activation functions, the weights for the isometricity enforcing
penalties, and the value of the parameter $\alpha$ used in 
the construction of the concave hull during the determination of
the disk boundaries. However, with the specific choices reported
here, the algorithm was found to be remarkably robust and worked
without modifications for all the examples presented in this paper
and in many other applications performed in ongoing work. We refer to Appendix \hyperref[s:Appendix]{A} for another test case that demonstrates how the algorithm handles shapes with sharp kinks. This gives
us confidence in the robustness of the method.

It is, of course, possible that for disks undergoing significantly
different types of motion/deformation some of our choices have to be
adjusted. If so, we recommend repeating the validation described
in \S \ref{sect:validation} with suitably modified synthetic data
sets. A visual inspection of the data produced by the fully-trained
encoder, as shown in Figure \ref{fig:boundary_determination}, will be helpful
to guide the choice of the radius of the sphere/disk used to determine
the boundary of the disk.

 Once the parameters have been adjusted so that representative
synthetic shapes can be reconstructed reliably, it still makes sense
to perform regular sanity checks on the shapes reconstructed
from actual experimental data for which no gold-standard exists.
We tend to check for spurious connections between adjacent sheets of
material and always monitor the conservation of the disk's area.

We implemented the algorithm in Mathematica 14.3, and the full, well-documented notebook
is openly available \cite{github}.
We stress that the main components of the algorithm are also available in
many other languages and open-source frameworks/libraries, so a
re-implementation should be straightforward. 

The low-cost data acquisition method used in our studies was perfectly
adequate for the relatively slowly moving elastic sheets considered in our
sedimentation experiments. Faster scanning methods with higher resolution
may be required to image elastic particles in high-speed flows. However,
as long as the imaging method is capable of describing the time-evolving
position and shape of the deforming sheet in a hypercloud, the algorithm
described in section~\ref{s:Reconstruction} is sufficiently generic that it can be used for any
such data sets.

\acknowledgements{This work was supported via EPSRC grants EP/P026044/1 and EP/T008725/1 and an EPSRC DTP studentship (TM).}

The data that support the findings of this article are openly available \cite{github}, embargo periods may apply.

\section*{Appendix A: Zigzag origami reconstruction}
\label{s:Appendix}

\begin{figure}[h]
    \centering
    \includegraphics[width=\textwidth]{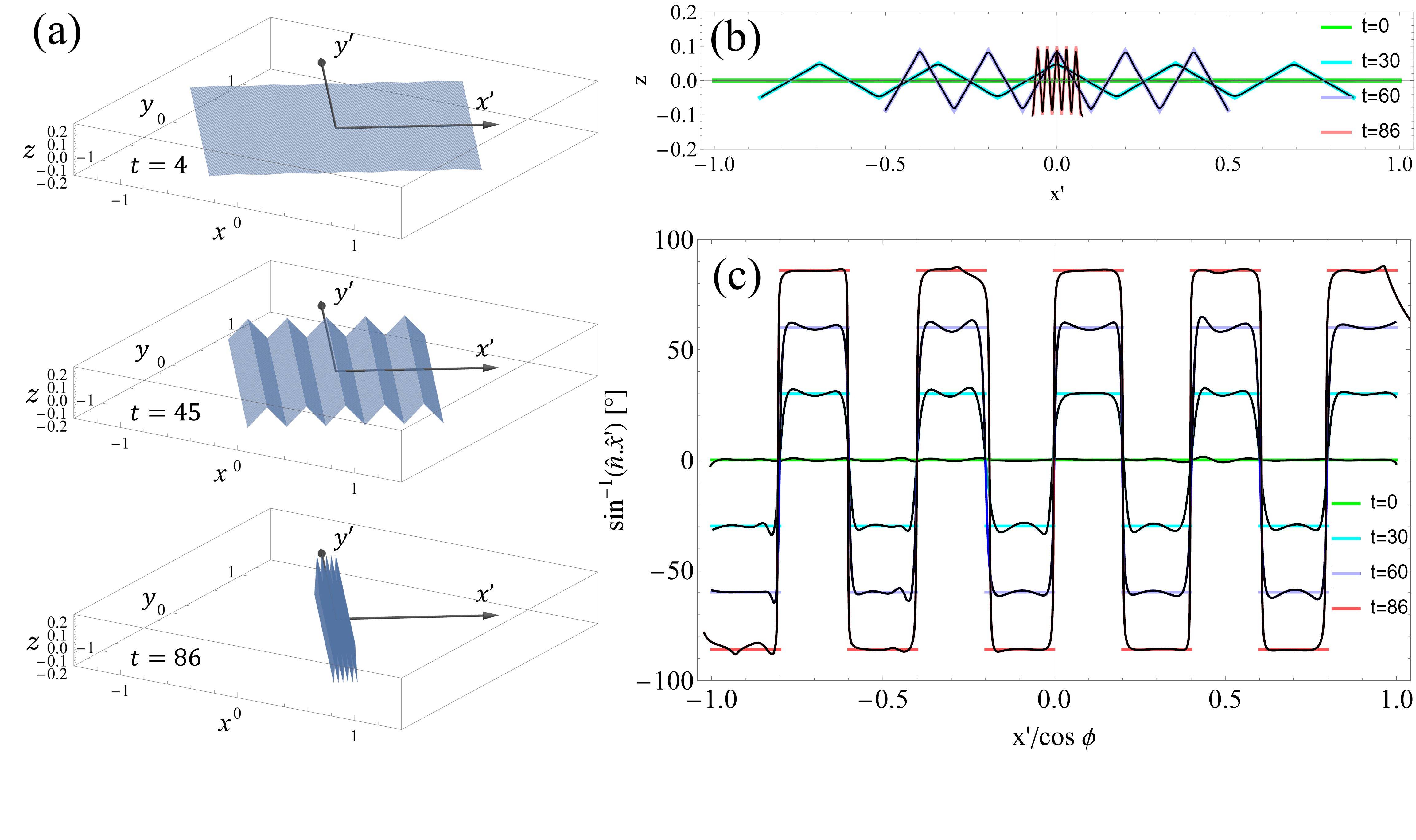}
    \caption{(a) Three selected time slices of the folding zigzag, whose horizontal axes $(x',y')$ rotated by $35\degree$ with respect to the $(x,y)$ axes. (b) Cross-section $\hat{y}'=0$ of the reconstructed surface (thin black lines) plotted on top of the ground truth for reference (thick coloured lines), which shows how the reconstruction rounds sharp corners. (c) Local inclination profile of the reconstructed surface (thin black lines) plotted against the ground truth (thick coloured lines).}
    \label{fig:ZigZag}
\end{figure}

In this appendix we present one further test case to demonstrate how the autoencoder handles shapes with sharp ridges. Given that our reconstruction algorithm approximates the shape by a smooth manifold, this provides a useful benchmark for the autoencoder's ability to handle geometrical features on short lengthscales. For this test, we created a synthetic zigzag dataset inspired by recent interest in origami structures that can form spontaneously in graphene sheets \cite{GrapheneZigZag2016}. The zigzag starts as a square of half-width 1 at $t=0$. Figure~\ref{fig:ZigZag}(a) shows how it folds in time into a zigzag shape with ten segments: the fold angle $\phi$ between the vertical and the normal vector of any segment changes linearly with time ($\phi = t\degree$, where $0 \le t \le 89$). We used $160$ thousand equally distributed points in each time slice. The zigzag's horizontal axes $(x',y')$ are rotated relative to the $(x,y)$ axes by an arbitrarily chosen angle of $35\degree$. 

The reconstruction was performed using the MPED cost function, without any physics-informed penalties.  Figure~\ref{fig:ZigZag}(b) shows that the reconstructed $y'=0$ cross-sections (thin black lines) accurately capture the target zigzag (thick coloured lines) as the fold angle increases, and thus, that the autoencoder has successfully reconstructed each zigzag shape. The largest discrepancy can be seen to occur at the pleats which appear rounded in the reconstruction. Figure~\ref{fig:ZigZag}(c) shows the inclination profile of the zigzag plotted against a rescaled coordinate $x'/\cos \phi$. The inclination of the reconstructed dataset (thin black lines) matches the target inclination at the centre of the segments (thick coloured lines) to within $3\degree$.  This demonstrates that the autoencoder is capable of reliably approximating shapes with sharp features.

\section*{Appendix B: Estimation of the timescales of elastic relaxation and sedimentation of the disk}
\label{s:AppendixB}
\textcolor{NewText}{
Here we estimate the timescale over which an elastic disk of bending
stiffness $D$ that is immersed in a viscous fluid of viscosity $\mu$
relaxes towards its undeformed planar equilibrium shape when it is
released from a configuration where it is bent into a U-shape. Denoting
the maximum out-of-plane displacement of the U-bent disk by $w \ll R$, the
disk's curvature is given by $\kappa \simeq w/R^2 \simeq \theta/R$,
where $\theta$ is a measure of the average angle by which each half
of the bent disk has been rotated out of the undeformed plane. The
bending moment on the disk's centreline is given by $T_b \simeq  \kappa D R \simeq \theta D$. While the disk relaxes to its undeformed configuration
its two halves rotate through the fluid with an average angular velocity
of $\simeq d\theta/dt$. This induces a resisting hydrodynamic torque
$T_h \simeq \mu R^3 \ d\theta/dt$ about the disk's centreline.
Equating the two torques then yields $D/(\mu R^3) \ \theta \simeq d\theta/dt$,
implying that the relaxation back to the undeformed state occurs over a
timescale
\begin{equation}
\mathcal{T}_{\rm relax} = \frac{\mu R^3}{D} \simeq \mathcal{O}(10^2{\rm s}).
\end{equation}
This order-of-magnitude estimate is consistent with the timescale observed in Figure \ref{fig:BendingEnergy} for the final stages
of the disk's relaxation. (Note that immediately following its release from
its strongly crumpled initial shape the disk deforms much more quickly.)
\\
The time it takes for a disk to traverse a single stationary light sheet can be estimated by dividing the disk diameter by the theoretical expression for the velocity of an edge-on fall of a flat circular disk \cite{Happel1983}:
\begin{equation}
    \mathcal{T}_{\rm translate}\simeq\frac{64\mu}{3\pi g b \Delta\rho}\simeq100~{\rm s}
 \end{equation}
 These estimates show that in our experiments sedimentation and deformation of the disk occur on similar timescales. Thus, the static mode of reconstruction introduced in \S \ref{s:ProjectorOpticsPolicy} cannot resolve the detail of the relaxation dynamics, which justifies the use of the dynamic mode in \S \ref{s:results} (see Appendix \hyperref[AppendixC]{C} for more detailed analysis).
}

\section*{Appendix C: Reconstruction at low scanning speeds}
\label{AppendixC}

\begin{figure}
    \centering
    \includegraphics[width=\linewidth]{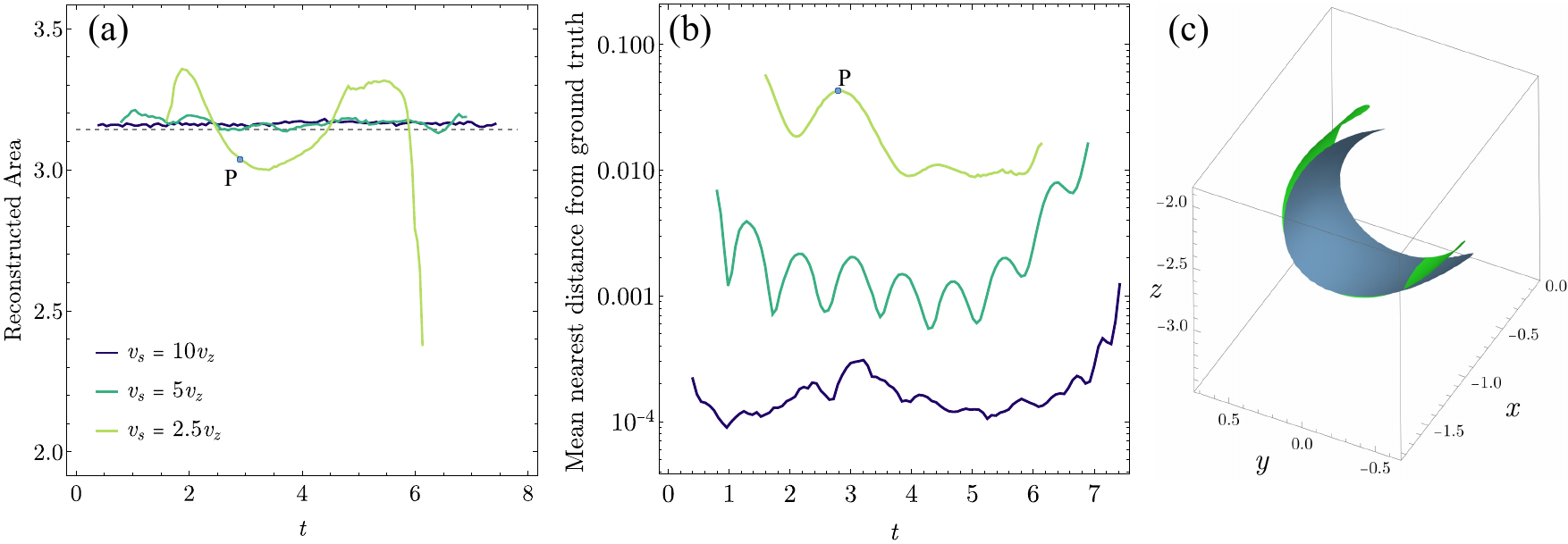}
    \caption{\label{fig:ScanningSpeedVariation}\textcolor{NewText}{(a) Time-traces of the area of the reconstructed disk from the synthetic dataset of the cylindrically-deforming disk introduced in  \S\ref{sect:validation} and shown in Figure \ref{fig:synthetic_disk_plot}(a), scanned at varying scanning speeds $v_s$ of 10 (dark blue line), 5 (green) and 2.5  (yellow) times the speed of the disk's vertical descent $v_z=1$. For reference, the true area $A=\pi$ of the disk is plotted as the dashed line. (b) Corresponding error of the reconstruction quantified by the mean of the nearest distances between all the sampled output points and the ground truth. (c) Comparison between the ground truth (green) and the output (blue) of one of the  worst-performing time slices $t=2.8$ of the dataset scanned at $v_s=2.5v_z$, labelled P in (a) and (b) showing that the reconstruction failed to correctly interpolate between the scans and the disk appears to be rotated relative to the ground tuth.}}
    \label{fig:placeholder}
\end{figure}

\textcolor{NewText}{In \S\ref{s:ProjectorOpticsPolicy} we stated that the details of the deformation or reorientation may be lost if the object is scanned slowly compared to the timescales of the deformation or reorientation. Here, we quantify this loss of accuracy of our reconstruction by producing, two additional synthetic datasets of the same simulated cylindrically-deforming disk introduced in \S\ref{sect:validation} and shown in Figure \ref{fig:synthetic_disk_plot}(a) but scanned at the speed 5 times and 2.5 times the speed of the disk's vertical descent $v_z=1$. We also show the original dataset scanned at the speed $v_s=10v_z$. 
The precise details of our arbitrarily imposed motion and deformation are as follows: the geometric centre of the disk follows $(x_{\rm cent}(t), y_{\rm cent}(t), z_{\rm cent}(t))=(\cos(t),\sin(t),-t)$ for $0<t<8$, the orientation is described by the nautical angles $\alpha, \beta,\gamma$ corresponding to three intrinsic rotations in the following order: rotation about $x$-axis by $\gamma=0.3t$ radians followed by a rotation about the (new) $y'$-axis by $\beta=0.2t$ radians followed by a rotation about the (new) $z''$-axis by $\alpha=0.1t$ radians. The disk and the radius of curvature changes as $R_c(t)=1+0.5\cos(t)$. 
Therefore, the timescales of most types of motion/deformation are of the order of $\sim 1$, while the duration of each scan in the original dataset is $\mathcal{T}=0.4$.
Reducing the speed of scanning by a factor of two (four) results in twice (four times) longer time between scans. As a result, the number of complete scans is reduced from the original twenty to ten  (five) and thus the datasets are undersampled.
\\ \\
Figure \ref{fig:ScanningSpeedVariation}(a) shows the effect this undersampling on the reconstructed area of the disk. The deviation from the true area of the disk $A=\pi$ (dashed line) increases the slower the scanning speed: for $v_s=10v_z$, the area fluctuates with standard deviation of $0.5\%$, for $v_s=5v_z$ the fluctuations rise to $1.7\%$ but for $v_s=2.5v_z$ the area fluctuates by $\sim 16\%$. In Figure \ref{fig:ScanningSpeedVariation}b we plot for each time instance the mean distance between the sampled output points and the corresponding nearest points on the ground truth. For the original dataset $v_s=10v_z$  (dark blue line) this distance is of the order of $\sim10^{-4}$, consistent with the value of ${\rm MPED}=1.7\times10^{-4}$ obtained during training. For slower scanning speeds, MPED values remained similar (${\rm MPED}=1.4\times10^{-4}$ and $2.6\times10^{-4}$ for $v_s=5v_z$ and $v_s=2.5v_z$, respectively), but the deviations from the ground truth are one or two orders of magnitude larger ($\sim 10^{-3}$ and $\sim 10^{-2}$, respectively). This indicates that while the reconstruction accurately matches any individual data point $(x_{ij},y_{ij},z_{ij},t_{ij})$ from the hypercloud, the interpolated regions of the sheet are distorted. One of the worst-performing reconstructions, from the time point $t=2.8$ of the slowest scanning speed (labeled P in Figs. \ref{fig:ScanningSpeedVariation}(a)(b)) is shown in Figure \ref{fig:ScanningSpeedVariation}(c), where the output surface (blue) appears rotated relative to the ground truth (green). }

%

\end{document}